\documentclass[fleqn,usenatbib]{mnras}

\usepackage{newtxtext,newtxmath}
\usepackage{setspace}

\usepackage[T1]{fontenc}

\DeclareRobustCommand{\VAN}[3]{#2}
\let\VANthebibliography\thebibliography
\def\thebibliography{\DeclareRobustCommand{\VAN}[3]{##3}\VANthebibliography}

\usepackage[utf8]{inputenc}
\usepackage[english]{babel}
\usepackage{listings}

\usepackage{graphicx}	

\usepackage[usenames,dvipsnames]{xcolor}
\usepackage{ulem}
\usepackage{hyperref}
\usepackage{subfloat}

\title[EGIPS catalogue]{The Edge-on Galaxies in the Pan-STARRS survey (EGIPS)}

\author[Makarov et al.]{
D.\ Makarov$^{1}$,
S.\ Savchenko$^{2,1}$, 
A.\ Mosenkov$^{3,4}$, 
D.\ Bizyaev$^{5,6,1}$, 
V.\ Reshetnikov$^{2,1}$,
A.\ Antipova$^{1}$, 
\newauthor
I.\ Tikhonenko$^{2}$, 
P.\ Usachev$^{1,2}$, 
S.\ Borisov$^{7,6,1}$, 
L.\ Makarova$^{1}$,
S.\ Kautsch$^{8}$, 
A.\ Marchuk$^{2,4}$, 
E.\ Rubtsov$^{6}$
\\
$^{1}$Special Astrophysical Observatory, Russian Academy of Sciences, Nizhnii Arkhyz, 369167 Russia\\
$^{2}$St.\ Petersburg State University, 7/9 Universitetskaya nab., St.\ Petersburg, 199034 Russia\\
$^{3}$Department of Physics and Astronomy, N283 ESC, Brigham Young University, Provo, UT 84602, USA\\
$^{4}$Pulkovo Observatory, Russian Academy of Sciences, St.\ Petersburg, 196140 Russia\\
$^{5}$Apache Point Observatory, New Mexico State University, Sunspot, New Mexico, USA \\
$^{6}$Sternberg Astronomical Institute, M.V. Lomonosov Moscow State University, Universitetsky prospect 13, Moscow, 119234 Russia\\
$^{7}$Department of Astronomy, University of Geneva, Chemin Pegasi 51, 1290 Versoix, Switzerland\\
$^{8}$Nova Southeastern University, Fort Lauderdale, FL 33314, USA\\
}

\date{Accepted XXX. Received YYY; in original form ZZZ}

\pubyear{2021}

\begin{document}
\label{firstpage}
\pagerange{\pageref{firstpage}--\pageref{lastpage}}
\maketitle

\begin{abstract}
We present a catalogue of 16,551 edge-on galaxies created using the public DR2 data of the Pan-STARRS survey.
The catalogue covers the three quarters of the sky above $\mathrm{Dec.}=-30\degr$.
The galaxies were selected using a convolutional neural network,
trained on a sample of edge-on galaxies identified earlier in the SDSS survey.
This approach allows us to dramatically improve the quality of the candidate selection
and perform a thorough visual inspection in a reasonable amount of time.
The catalogue provides homogeneous information on astrometry, \textsc{SExtractor} photometry, 
and non-parametric morphological statistics of the galaxies.
The photometry is reliably for objects in the 13.8--17.4 $r$-band magnitude range.
According to the HyperLeda database, redshifts are known for about 63 percent of the galaxies in the catalogue.
Our sample is well separated into the red sequence and blue cloud galaxy populations.
The edge-on galaxies of the red sequence are systematically $\Delta(g-i)\approx0.1$~mag redder 
than galaxies oriented at an arbitrary angle to the observer.
We found a variation of the galaxy thickness with the galaxy colour.
The red sequence galaxies are thicker than the galaxies of the blue cloud.
In the blue cloud, on average, thinner galaxies turn out to be bluer.
In the future, based on this catalogue it is intended to explore 
the three-dimensional structure of galaxies of different morphologies, 
as well as to study the scaling relations for discs and bulges.
\end{abstract}




\section{Introduction}

Disc galaxies inclined at nearly $90\degr$ to the line-of-sight and often called edge-on galaxies 
are the only extragalactic objects whose vertical structure can be studied directly.
Since early studies by \citet{1978ApJ...223L..63K,1979ApJ...234..829B,1981A&A....95..105V}, 
many important results have been obtained on the vertical distribution of matter in the disc, 
bulge, and halo of the edge-on galaxies.
For example, it was shown that the existence of superthin galaxies 
with an axial ratio $a/b>10$ 
is only possible in the presence of a massive dark halo surrounding the disc \citep{1991SvAL...17..374Z}.
The combination of photometric and kinematic data allows researchers 
to estimate the parameters of the dark halo \citep[see e.g. ][]{2021ApJ...914..104B}.
Further analysis of the rotation curves of superthin galaxies indicates the presence
of a compact dark matter halo whose pseudo-isothermal dark matter core radius is smaller than two radial disc scale-lengths
\citep{2017MNRAS.466.3753B,2018MNRAS.479.5686K,2021ApJ...914..104B}.
Knowing the disc thickness allows us 
to impose a limit on the halo mass~\citep{2006AstL...32..649S,2010AN....331..731K}.

Flat galaxies with $a/b>7$ have proven to be a good tool 
for studying bulk motions of galaxies in the Universe \citep{2000BSAO...50....5K,2003A&A...407..889K}.
Their edge-on orientation eliminates one of the biggest uncertainties in the Tully-Fisher relation 
-- the inclination correction.
It allows us to estimate the distance modulus to flat galaxies with an accuracy of 0.32~mag \citep{2018MNRAS.479.3373M}.
Additionally, bulgeless flat galaxies (also called simple discs)
are an ideal tool to verify different galaxy formation scenarios 
that challenge the existing evolution theories of disc survival in a merger-dominated Universe \citep{2009PASP..121.1297K}.

Considering large, uniformly selected samples of edge-on disc galaxies
allows us to study the vertical structure statistically, which
in turn helps understand a variety of external and internal processes \citep{2004ARA&A..42..603K,2015HiA....16..316K}
that play a role in galaxy evolution.
This includes the shapes of the different bulge types 
\citep[classical versus boxy, see][]{2005RMxAC..23..101K} 
and their relationships with bars and other galactic components. 
The vertical structure can also test predictions of competing theories 
of the thick disc formation such as an in-situ formation due to the dissolution of giant star formation regions \citep[e.g.][]{2002MNRAS.330..707K}, 
disc heating by mergers \citep[e.g.][]{1993ApJ...403...74Q}, 
formation of discs by accretion of stars from satellites \citep[e.g.][]{2003ApJ...597...21A},
growing up discs via instabilities from material of the thin disc \citep[e.g.][]{1967ApJ...150..461B} 
and dynamical scattering \citep{1985ApJ...290...75V}.

Only few catalogues focus on uniformly selected edge-on galaxies and provide their physical properties. 
The classic Flat Galaxies Catalogue~\citep{1993AN....314...97K} and 
its updated version, 
the Revised Flat Galaxy Catalogue~\citep[RFGC,][]{1999BSAO...47....5K}, contain 4236 thin spiral galaxies 
with a blue diameter $a\ge40$~arcsec and a blue major-to-minor axis ratio $a/b\ge7$ found over the whole sky
by systematic visual inspection of all the prints of the Palomar Observatory Sky Survey (POSS-I) 
and the ESO/SERC sky survey in the blue and red colours. 
The edge-on galaxies in the Sloan Digital Sky Survey Data Release~1 \citep[SDSS,][]{2003AJ....126.2081A} 
and 6 \citep{2008ApJS..175..297A} were catalogued and automatically classified 
into various morphological Hubble types by \citet{2006A&A...445..765K} and \citet{2009AN....330..100K}, respectively.
Finally, the catalogue of edge-on disc galaxies \citep[EGIS,][]{2014ApJ...787...24B}
contains 5747 genuinely edge-on galaxies visually inspected 
after automatic selection from the SDSS DR7 \citep{2009ApJS..182..543A}.

In this article we introduce our catalogue of 16,551 Edge-on Galaxies in the Pan-STARRS survey (EGIPS).
The galaxies were identified over the $3/4$ of the entire sky with $\mathrm{Dec.}>-30\degr$ 
in the Pan-STARRS DR2 images \citep{2016arXiv161205560C,2020ApJS..251....7F}.
The EGIPS candidates were selected by a Convolutional Neural Network (CNN) 
followed by an accurate visual inspection made by well-experienced professional astronomers.
Our catalogue provides information on positions, photometric and morphological parameters of the edge-on galaxies, and
cross-identification with the HyperLeda~\citep{2014A&A...570A..13M} 
and RCSED\footnote{\url{http://rcsed.sai.msu.ru/}}~\citep{2017ApJS..228...14C} databases.
The public access to the EGIPS catalogue is supported by the Edge-on Galaxy Database\footnote{\url{https://www.sao.ru/edgeon/}} \citep{2021AstBu..76..218M}.

\section{Candidate selection from the Pan-STARRS images}

\label{sec:CandidateSelection}

Our selection of edge-on galaxies is performed using imaging from the Panoramic Survey Telescope and Rapid Response System~\citep[Pan-STARRS,][]{2016arXiv161205560C}.
The Pan-STARRS survey is carried out in five ($g$, $r$, $i$, $z$, $y$) broadband filters 
using the 1.8-meter telescope located at the Haleakala Observatory (Hawaii, US) 
and equipped with a 1.4 Gigapixel camera.
The survey covers the whole Northern and part of the Southern hemisphere down to $\mathrm{Dec.}=-30\degr$, 
with a typical seeing of 1.31, 1.19 and 1.11~arcsec \citep{2020ApJS..251....6M}
and a photometric limit of 23.3, 23.2 and 23.1~mag in the $g$, $r$, $i$ bands, respectively.
The access to the publicly available image and object catalogue archives \citep{2020ApJS..251....7F} 
is provided by the Space Telescope Science Institute (STScI).
The Pan-STARRS images\footnote{\url{https://outerspace.stsci.edu/display/PANSTARRS/PS1+Sky+tessellation+patterns}} 
are interpolated on a regular grid of $4\degr\times4\degr$ projection cells covering the sky,
which in turn are divided into $10\times10$ skycells.
Each skycell is $0.4\degr\times0.4\degr$ with a pixel size of 0.25~arcsec.
The name format for a skycell image is \texttt{skycell.nnnn.0yx},
where \texttt{nnnn} is the projection cell number
and \texttt{0yx} indicates the location of the skycell in the projection cell.
The projection cell number is in the range from 635 to 2643.
The coordinates \texttt{y} and \texttt{x} vary from 0 to 9 indicating the sub-cell within a projection cell.

Our first attempt to select edge-on galaxy candidates
using an automatic catalogue of extended sources 
generated by the Pan-STARRS pipeline~\citep{2020ApJS..251....5M} was unsuccessful.
We used the RFGC catalogue~\citep{1999BSAO...47....5K} as a reference sample to find the best criteria 
for selecting flat galaxies in the Pan-STARRS object catalogue.
Unfortunately, our analysis did not show a good correlation 
between the properties of the RFGC galaxies and the corresponding objects in the Pan-STARRS catalogue.
Most likely, extended and highly elongated objects require 
a fine tuning of parameters for their successful selection.
Particularly, universal object finding algorithms break flat galaxies into pieces, 
incorrectly determine the size and ellipticity, and/or even do not detect edge-on galaxies at all.
In order to provide an acceptable loss rate of less than 20 per cent of the target galaxies, 
we had to make the selection criteria too loose,
which, in turn, led to a catastrophic increase in the number of false candidates in the sample.
As a result, the good-to-bad ratio became 1 to 500 or even worse.
This failure prompted a search for a new approach.

Next, for selecting edge-on galaxies, we decided to use the artificial neural network (ANN) methodology.
A detailed description of the ANN architecture and the training process used in this study 
can be found in an accompanying article (Savchenko et al.\ in prep.). 
Here we briefly outline the basic procedure.
Typically, a convolutional neural network \cite[CNN,][]{LeCun1989, LeCun1990} is used 
for providing an efficient pattern recognition in an image analysis.
CNN is a special type of neural network that takes into account the spatial relations between image pixels.
After a series of experiments with CNN architectures, we settled on the following option.
Our classification system includes three blocks. 
Each block consists of two convolutional layers with normalisation \citep{Ioffe2015}, max pooling \citep{Boureau2010},
and a 30 per cent dropout \citep{Srivastava2014},
where the number of convolutions increases in each convolution block.
This is followed by a fully connected layer for decision making.
At the end, there is an output layer with two neurons for the classification of edge-on galaxies and other objects.
The total number of trainable parameters of the network is 206,894.
This scheme was implemented using the \textsc{tensorflow}\footnote{\url{https://www.tensorflow.org/}} package 
in the \textsc{python}\footnote{\url{https://www.python.org/}} programming language.

For our neural network training we use the sample consisting of 5747 edge-on galaxies 
from the EGIS catalogue, 
of which 80 per cent are the training sample
and the remaining 20 per cent are test objects.
We made a training sample of negative examples to discriminate non-edge-on galaxies.
It consists of 54,000 galaxies taken from the HyperLeda database
with an apparent major-to-minor axis ratio less than 4, $\mathtt{logr25}<0.6$,
in order to cut off highly inclined galaxies,
and a major diameter between 0.1 and 0.5~arcmin, $0<\mathtt{logd25}<0.7$, 
to exclude too small and over-smoothed objects, as well as extremely extended galaxies, 
that could suffer from the sky subtraction pipeline in the Pan-STARRS survey.
To ensure that we do not have some edge-on galaxies among our negative examples, 
we also imposed the following restrictions on the Galaxy Zoo~\citep{2008MNRAS.389.1179L} vote fractions 
for all the objects in the negative subsample: 
$\mathtt{P\_EDGE}$ (an edge-on galaxy) is less than 0.1 
and at least one of the following parameters $\mathtt{P\_EL}$ (an elliptical galaxy), 
$\mathtt{P\_CW}$ (a spiral galaxy winded clockwise) 
or $\mathtt{P\_ACW}$ (a spiral galaxy winded counterclockwise) is greater than 0.1.
The negative examples were supplemented by a random sample of stars and also by empty background fields.
Unfortunately, these positive and negative samples are quite small for an effective deep learning by our CNN.
To solve this problem, we used data augmentation by applying the following procedure: 
adding noise to the images, 
vertical and horizontal flipping of the images, 
rotation of the images at a random angle,
zooming-in/out of the images by a random factor up to 25 per cent.
These steps allowed us to increase the training sample to up to 300,000 objects.
After 50 epochs of learning, our CNN reached an accuracy of classification better than 99 per cent with the precision (the number of true positive detections divided by the total number of true positives and false positives) and recall (the number of true positives divided by the sum of true positives and false negatives) values equal to 0.993 and 0.991, respectively.
To assess the robustness of our selection method, we tested a system with 3027 RFGC galaxies that were not included in our training sample.
Only 20 of them were misclassified, giving an error rate smaller than 1 per cent.
We independently trained five CNN models to improve the quality of the classification.

We download the $g$, $r$ and $i$-band images of each Pan-STARRS skycell.
Then, we detect all objects in the $r$ band with the \texttt{detect\_source} function 
of the \textsc{photutils} library \citep{2020zndo...4044744B} in \textsc{python}.
To meet our detection threshold, an object must be at least 48 pixels in length
with at least 15 connected pixels above 4 standard deviations of the background level.
The 48 pixel size ($=12$~arcsec) was chosen to reliably measure the galaxy thickness
even for very thin galaxies, given the typical seeing of 1.19~arcsec in the $r$ band.
The image of each object is extracted from the Pan-STARRS skycells in all 3 bands 
and scaled to a $48\times48$ pixel size.
The stacked 3D array of the $g$, $r$ and $i$ images is fed into our neural network.
An object is considered as a good edge-on galaxy candidate
if the majority of CNN models (at least three out of five) 
have voted positively (i.e.\ the probability is greater than 0.5)
for the edge-on class.
We processed all the $\sim200,000$ skycells of the Pan-STARRS archive
in the above described manner
and selected 26,719 edge-on galaxy candidates.

\begin{subfigures}
\begin{figure}
\centering
\includegraphics[width=0.23\textwidth]{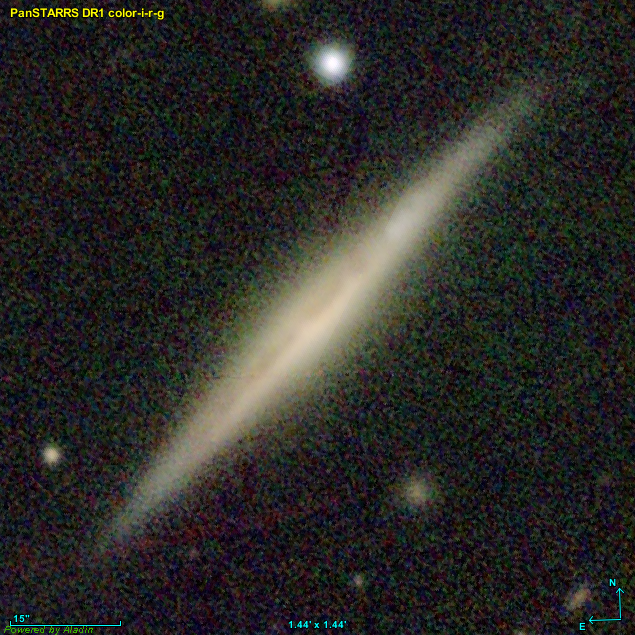}
\includegraphics[width=0.23\textwidth]{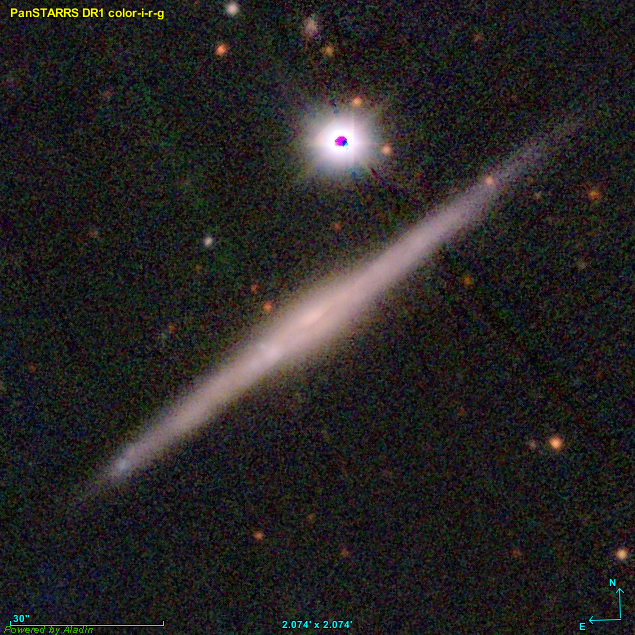}
\includegraphics[width=0.23\textwidth]{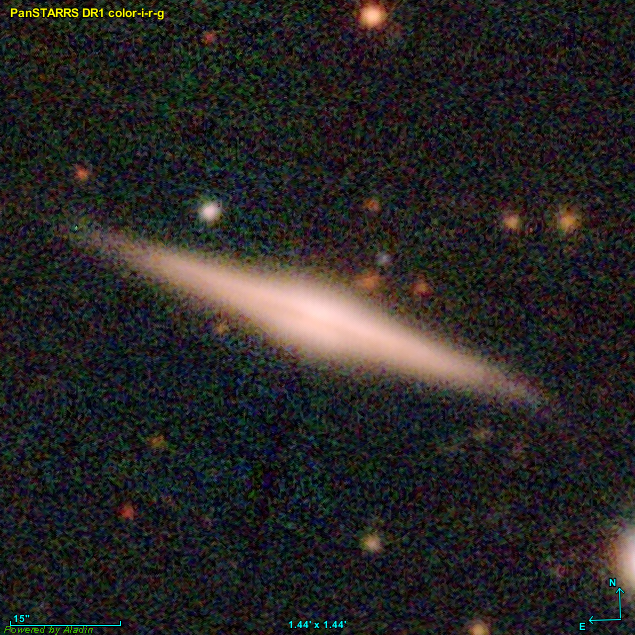}
\includegraphics[width=0.23\textwidth]{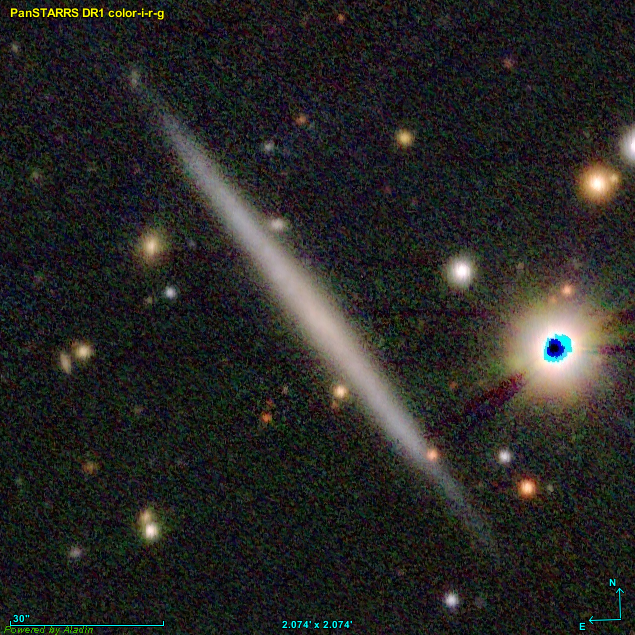}
\caption{
Examples of objects classified as truly edge-on galaxies.
}
\label{fig:exampleGood}
\end{figure}

\begin{figure}
\centering
\includegraphics[width=0.23\textwidth]{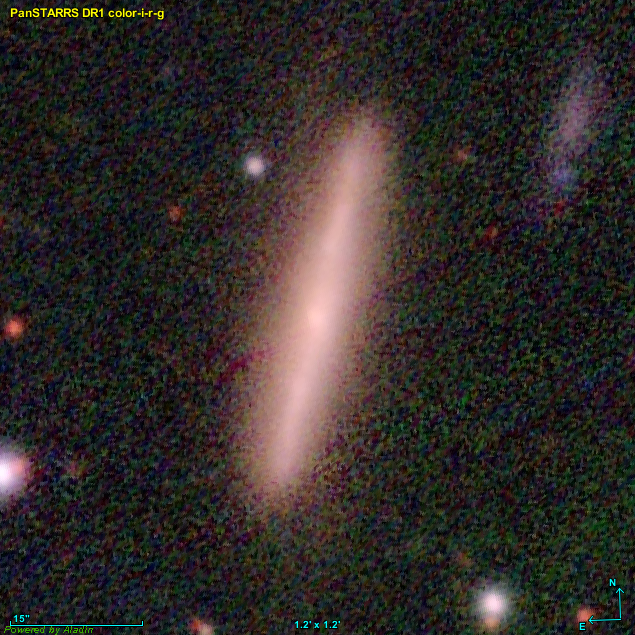}
\includegraphics[width=0.23\textwidth]{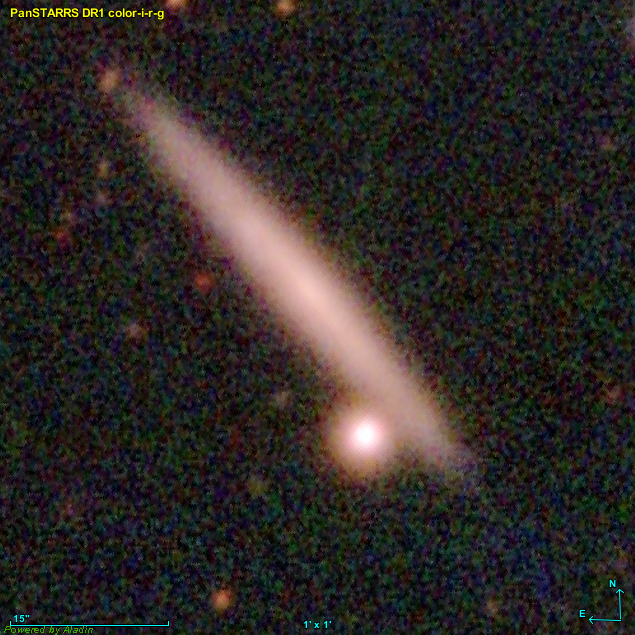}
\includegraphics[width=0.23\textwidth]{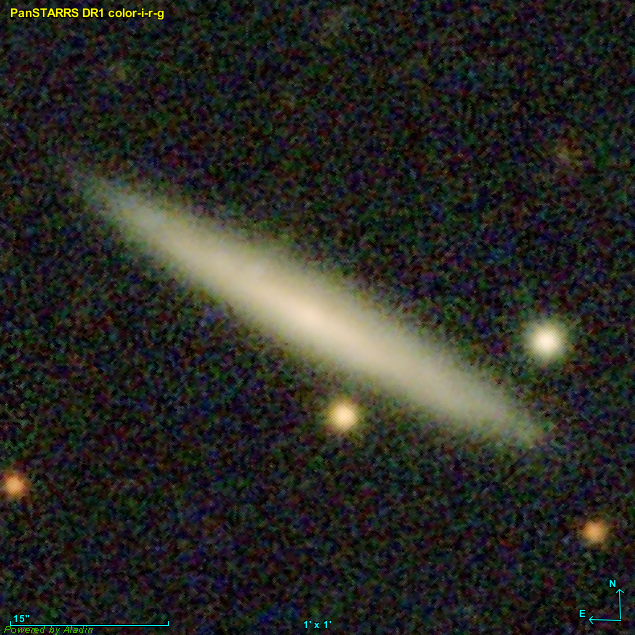}
\includegraphics[width=0.23\textwidth]{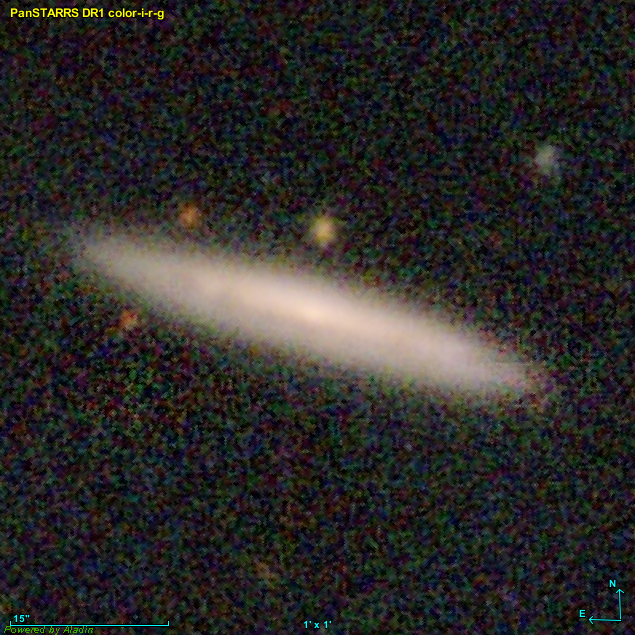}
\caption{
Examples of galaxies classified as highly inclined and acceptable for including in our catalogue of edge-on galaxies.
}
\label{fig:exampleAcceptable}
\end{figure}

\begin{figure}
\centering
\includegraphics[width=0.23\textwidth]{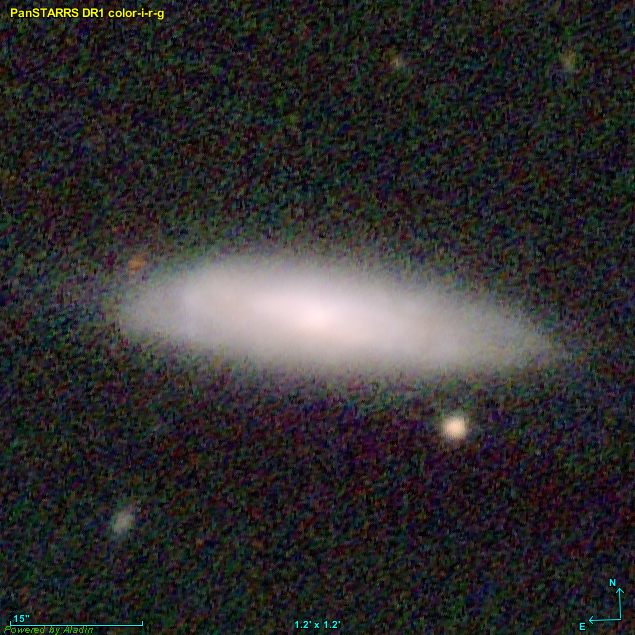}
\includegraphics[width=0.23\textwidth]{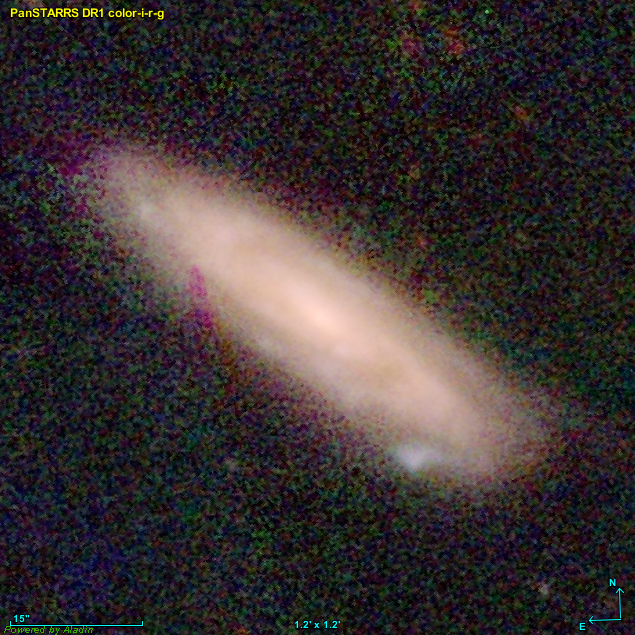}
\includegraphics[width=0.23\textwidth]{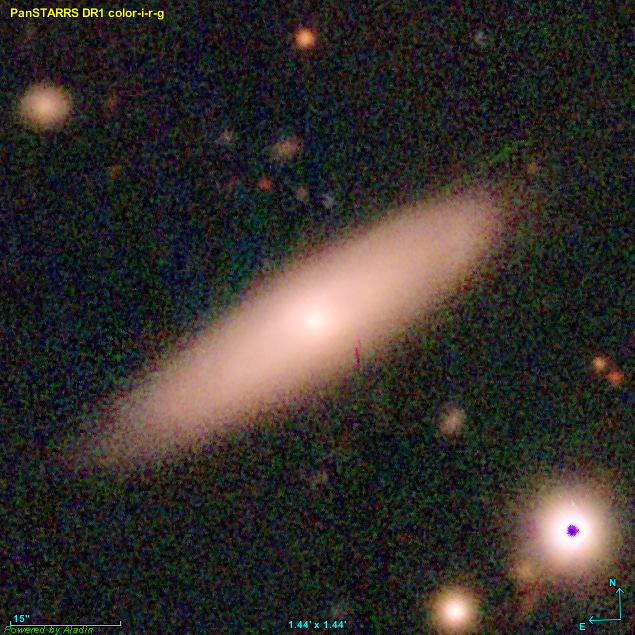}
\includegraphics[width=0.23\textwidth]{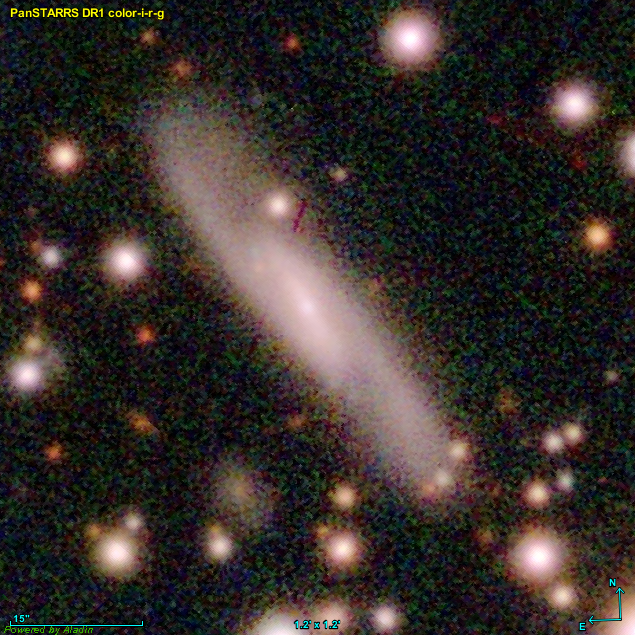}
\caption{
Examples of galaxies classified as unsuitable for our catalogue of edge-on galaxies.
}
\label{fig:exampleUnsuitable}
\end{figure}
\end{subfigures}

Despite of the significant improvement in the quality of our selection of edge-on galaxies using the ANN methodology, 
the final sample of candidates contains a significant number of objects 
that were falsely interpreted by the ANN as edge-on galaxies. 
Typically, misclassified objects are asterisms, image defects, artefacts from bright stars, 
and non-edge-on galaxies.

We performed a visual inspection of the candidates to reject different types of artefacts and 
cases of wrong classification.
During the inspection, we also estimated the proximity of the candidates to an edge-on orientation: 
\begin{itemize}
\item good candidates -- truly edge-on galaxies (see Fig.~\ref{fig:exampleGood});
\item acceptable candidates -- highly inclined galaxies, $i\gtrsim80\degr$ (Fig.~\ref{fig:exampleAcceptable});
\item unsuitable candidates -- genuine galaxies that do not satisfy the previous conditions (Fig.~\ref{fig:exampleUnsuitable}).
\end{itemize}
Our voting poll was organized on the Zooniverse\footnote{\url{https://www.zooniverse.org/}} web portal for citizen science.
At this stage, 12 well-experienced astronomers took part in the visual inspection 
and at least three participants examined each edge-on candidate.
Several examples of clear classes, according to the consensus of the classifiers, are shown in Figures~\ref{fig:exampleGood}, \ref{fig:exampleAcceptable} and \ref{fig:exampleUnsuitable}.

Fortunately, asterisms, image artefacts, and bright stars can be detected in the images very easily and reliably.
In each case, the participants showed a complete agreement on the classification of such wrongly classified cases.
In total, we excluded 3992 such objects from a further consideration.

The estimate of the proximity of galaxies to an edge-on orientation caused a certain spread in opinions.
Some voters tended to apply very strict criteria for good candidates, while some others turned out to be more liberal,
despite of the fact that all participants in the classification followed the same instructions.
Nevertheless, this information proved to be extremely important for the creation of the final version of the catalogue.
To be included in the catalogue, a candidate must be marked as `good' (truly edge-on) by at least one classifier,
or must have no more than 70 per cent of `unsuitable' votes.
Finally, we formed a catalogue of 16,551 nearly or purely edge-on galaxies.

The comparison of the final sample with the initial list of 26,719 candidates 
(22,727 of which are genuine galaxies)
shows that about 10 thousand objects did not pass the visual inspection.
Thus, the use of the CNN made it possible to reduce the number of candidates to an acceptable level 
with a fraction of edge-on galaxies of over 60 per cent.
This is a significant progress compared to our first attempt to select extended sources from the Pan-STARRS pipeline.

\section{Photometry}
\label{sec:Photometry}

We performed \textsc{SExtractor}~\citep{1996A&AS..117..393B} photometry 
using $g$, $r$, $i$, $z$, $y$ Pan-STARRS imaging
for all the 22,727 candidates remaining in the sample after excluding wrongly classified objects that are not galaxies.
Unfortunately, it is impossible to choose the \textsc{SExtractor} parameters for our automatic photometry,
so that they work equally well in all filters and for all galaxies.
\textsc{SExtractor} has two parameters, \texttt{deblend\_nthresh} and \texttt{deblend\_mincount}, 
which control the process of object identification in images.
The \texttt{deblend\_nthresh} parameter specifies the number of levels in the flux, 
at which \textsc{SExtractor} searches for local minima between different objects to separate them.
The \texttt{deblend\_mincount} parameter sets the fraction of the flux that a local maximum must have in order to be selected as a separate object.
Most galaxies from our sample were processed well, but in some cases the object in question could be split into pieces.
Another reason for failure can be an object, usually a star, which overlaps with our target galaxy and makes its photometry hard to characterize.
In such cases, \textsc{SExtractor} often yields a problematic output in several filters, so that it is quite easy to find such problematic objects 
by comparing their properties in different bands.
It turned out the coordinates and position angle are extremely sensitive to the object identification problems, 
as well as to the superimposition of stars and close galaxies. 
We mark the results of our photometry in a given band as suspicious 
if the position angle deviates from the median value for all filters by more than 2.5~degrees
or if the coordinates differ by more than a certain fraction of the corresponding major axis $a$ from the median position
(specifically, $>0.3 a_r$, $>0.4 a_i$ and $>0.6 a_g$).
To solve the segmentation problems, we performed a visual inspection of about 5500 in-doubt objects.
The user was able to vary the \texttt{deblend\_mincount} and \texttt{deblend\_nthresh} parameters 
to provide the best description of the galaxy in all filters.
In most cases, it was possible to find a good set of parameters for all filters, 
otherwise the problematic filter was flagged as unreliable.
Finally, all candidates for edge-on galaxies were reprocessed with \textsc{SExtractor} using the manually-chosen segmentation parameters if needed.

These steps allowed us to obtain the following properties of the galaxies based on the \textsc{SExtractor} photometry in the five Pan-STARRS bands:
the astrometry (the coordinates of the object's barycentre),
the basic shape parameters (the semi-major and semi-minor axes of the ellipse describing the object, 
the position angle, as well as the ellipticity and the elongation), and
the \citet{1980ApJS...43..305K} and \citet{1976ApJ...209L...1P} estimates of the total flux.

We had some doubts about the reliability of our photometry.
The Pan-STARRS sky subtraction pipeline overestimates the background brightness near very extended objects, 
which leads to artificial depressions of brightness around bright galaxies
and distorts their photometry.
Also the Pan-STARRS data reduction pipeline produces noticeable background ripples, 
that are clearly visible in large-scale images. 
This can affect results of photometry for very extended galaxies.
Background and foreground objects can also significantly distort the photometry. 
Automatic methods are not always able to take into account these cases properly.
Strongly elongated galaxies are also traditionally difficult cases for automatic processing.

\begin{figure}
\centering
\includegraphics[width=0.49\textwidth]{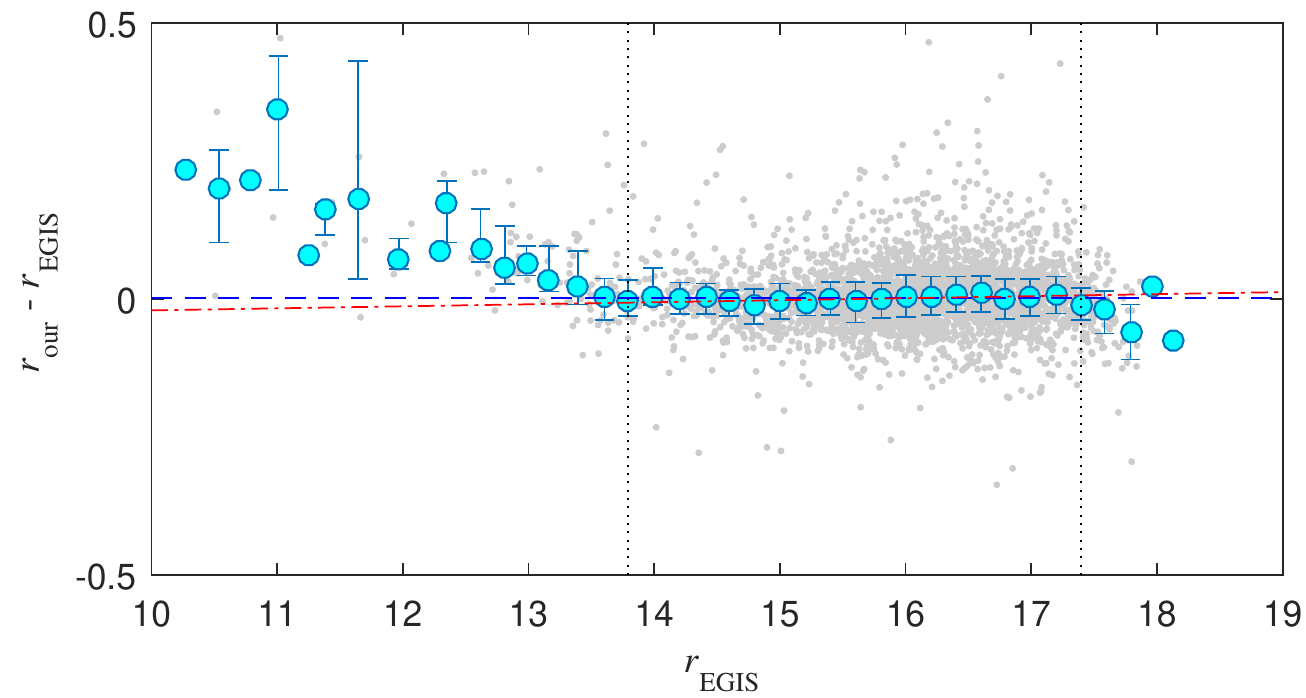}
\caption{
Gray dots show the difference between our Petrosian and the EGIS aperture magnitudes in the $r$ band.
Large cyan dots present the running median values within a window of 0.2 mag 
and their error bars correspond to the first and third quartiles.
The black dotted vertical lines indicate the agreement range.
The blue dashed line corresponds to the median value, 
while the red dash-dotted line represents a robust linear fitting for the region within the agreement range.
}
\label{fig:PS1SDSSPhotoComparison}
\end{figure}

\begin{table}
\centering
\caption{
Comparison between the Petrosian magnitudes (this work) 
and the EGIS aperture photometry in the different bands.
Our photometry was made in the Pan-STARRS photometric system,
while the EGIS data were processed in the SDSS system.
}
\begin{tabular}{cr@{--}lcccc}
\hline\hline
band &
\multicolumn{2}{c}{range} &
\% &
\multicolumn{3}{c}{$\Delta=m_{\rm our}-m_{\rm EGIS}$} 
    \\
    &
\multicolumn{2}{c}{EGIS} &
    &
\multicolumn{1}{c}{Median} & 
$\sigma$  & 
slope 
    \\
\hline
$g$ & 14.8 & 18.0 & 91 & $-0.032\phantom{0}$ & 0.067 & $+0.0106\pm0.0018$ \\
$r$ & 13.8 & 17.4 & 93 & $+0.0022$           & 0.048 & $+0.0037\pm0.0012$ \\
$i$ & 13.2 & 17.2 & 96 & $+0.025\phantom{0}$ & 0.044 & $-0.0016\pm0.0010$ \\
\hline\hline
\end{tabular}
\label{tab:PS1SDSSPhotoComparison}
\end{table}

Fortunately, we have the opportunity to test the quality of our automatic \textsc{SExtractor} photometry 
against a completely independent processing.
For all galaxies in the EGIS catalogue, \citet{2014ApJ...787...24B} performed
aperture photometry in the three $g$, $r$, $i$ SDSS bands using SDSS DR7 images.
The flux was measured inside an ellipse corresponding to the galaxy isophote at a signal-to-noise ratio $S/N = 2$ per pixel,
which is quite close to the total flux of the galaxy.
Most of the EGIS galaxies are included in our catalogue (see Section~\ref{sec:Comparison}).
However, since the published EGIS photometry is corrected for the foreground Galactic extinction
using the dust maps of \citet{1998ApJ...500..525S}, 
for a proper comparison we de-correct their values 
by adding the appropriate extinction correction to the EGIS magnitudes.
Fig.~\ref{fig:PS1SDSSPhotoComparison} illustrates the behaviour of the difference between our Petrosian 
and the EGIS aperture magnitudes, 
$\Delta=r_{\rm our}-r_{\rm EGIS}$, versus the EGIS estimates in the $r$ band.
The diagrams for the $g$ and $i$ bands look very similar, so we do not display them.

We find an extremely good agreement between our and EGIS magnitudes, especially in the $r$ and $i$ bands, 
despite of the known issues with the photometry of extended objects,
the difference between the Pan-STARRS and SDSS photometric systems,
and the difference in the methodologies (our Petrosian magnitudes versus the aperture photometry in EGIS).
The results of the comparison are gathered in Table~\ref{tab:PS1SDSSPhotoComparison}.
The first column indicates the filter,
the second column shows the range of the best agreement between our and EGIS photometries,
the 3rd column indicates the percentage of galaxies of the sample in the agreement region,
the 4rd and 5th columns give the median value of the difference and its scatter,
and the 6th column contains the slope of the robust linear regression with its error for the data inside the agreement region.
The absolute shift between the magnitudes determined by us and those from EGIS is not very relevant 
because it is affected by the difference in the methodologies and the photometric systems.
However, it is encouraging to see that this shift is quite small.
It is important that the scatter in the data is extremely small, less than 0.05~mag in the $r$ and $i$ bands.
We see a significant slope only in the case of the $g$ filter.
In the $r$ band, the slope has a significance at the level of three sigma, but with quite a small value.
In the $i$ band, the agreement between the photometries is excellent.
The slope is insignificant and the scatter is minimal.
Thus, in all 3 bands under consideration there is an agreement region with a width of about 3.5 mag without significant systematics,
where we can rely on our automatic photometry.
Outside the agreement region, we see the following expected trends.
The total flux of bright galaxies is systematically underestimated
due to the specifics of the sky subtraction procedure in Pan-STARRS.
At the faint end near our photometric limit
the statistics suffer from the Malmquist bias \citep{1922MeLuF.100....1M}.

In addition to the \textsc{SExtractor} photometry, we calculated non-parametric morphological statistics 
using the \textsc{statmorph} package~\citep{2019MNRAS.483.4140R}. 
To calculate the statistics, we used the same segmentation maps, 
that we obtained in our preparation of the photometry.
That allowed us to determine a number of important statistics: 
the relative distribution of the pixel flux values in a galaxy, also known as the Gini coefficient \citep{2004AJ....128..163L};
the second-order moment of the brightest 20 per cent of the galaxy's flux, $M_{20}$ \citep{2004AJ....128..163L};
the concentration, asymmetry and smoothness (CAS) indexes of the light distribution in a galaxy \citep{2003ApJS..147....1C} 
(see the description of these quantities in Section~\ref{sec:Catalogue}).

\section{Database and Catalogue}
\label{sec:Catalogue}

We present two data sets, 
namely the edge-on galaxy candidate catalogue\footnote{\url{https://www.sao.ru/edgeon/catalogs.php?cat=PS1candidate}} 
and the catalogue of edge-on galaxies\footnote{\url{https://www.sao.ru/edgeon/catalogs.php?cat=EGIPS}} in the Pan-STARRS survey (EGIPS). 
Both are published on-line in the Edge-on Galaxies Database\footnote{\url{https://www.sao.ru/edgeon/}} \citep{2021AstBu..76..218M}. 
The Edge-on Galaxies database was developed in order to systematize the information, 
simplify its use, and facilitate the data analysis.
In fact, visual classification and cross-identification of edge-on galaxy candidates 
was carried out with the aid of the interfaces of the database.
The database provides access to HyperLeda data for cross-identified objects,
as well as access to digitized astronomical surveys and catalogues 
using Aladin Sky Atlas\footnote{\url{https://aladin.u-strasbg.fr/}}~\citep{2000A&AS..143...33B,2014ASPC..485..277B}.

The catalogue of edge-on galaxy candidates
consists of a number of tables \citep[Section 2.4]{2021AstBu..76..218M}.
The list of candidates was prepared using the \textsc{photutils} library \citep{2020zndo...4044744B} 
as described in Section~\ref{sec:CandidateSelection}.
In total, it contains 22,727 genuine galaxies.
The table provides the candidate ID as a primary key, its coordinates, and preliminary photometric data.
As a candidate ID, we use a set of three numbers: 
the skycell identifier, which consists of the projection cell and sub-cell numbers in the Pan-STARRS sky tessellation 
(see Section~\ref{sec:CandidateSelection}),
and the internal candidate number in a given skycell field.
For example, due to a slight overlap of neighboring skycell images 
the galaxy EGIPS~J095635.9+203843 appears in the candidate catalogue two times 
under the names \texttt{1799\_016.0} (projcell=1799, subcell=016, candidate=0) and \texttt{1799\_017.0} (projcell=1799, subcell=017, candidate=0).
All candidates were cross-matched with objects from the HyperLeda database.
As a result, each of our candidates was assigned a PGC number.
This allows us to link the candidates with objects in other catalogues.
Also this step automatically identified duplicates among our candidates.
The PGC number is a convenient key for binding and combining the data.

The list of candidates is accompanied by a table collecting the votes of the five CNN models
and by a table of visual classifications.
For usability, this set of data tables is supported by dynamically generated auxiliary virtual tables
for calculating and organizing the classification statistics for each object.
The detailed photometry measurements made with \textsc{SExtractor}~\citep{1996A&AS..117..393B} 
and the non-parametric morphology parameters measured with \textsc{statmorph}~\citep{2019MNRAS.483.4140R} 
are listed in the respective tables.
In the framework of the project, we performed two runs of the \textsc{SExtractor} photometry. 
The second run was carried out after manual tuning of the \textsc{SExtractor} parameters 
(see Section~\ref{sec:Photometry}).
The results of the first and second runs are stored separately in two different tables.

The catalogue of 16,551 edge-on galaxies in the Pan-STARRS survey
is a subsample of candidates selected as the most likely edge-on galaxies after the visual inspection
as described in Section~\ref{sec:CandidateSelection}.
The main table of the catalogue lists the true edge-on galaxies and their coordinates.
We introduce new designations with the acronym EGIPS followed by the coordinates as recommended by the IAU\footnote{\url{http://cdsweb.u-strasbg.fr/Dic/iau-spec.html}}.
The tables of the visual classification, \textsc{SExtractor} photometry and \textsc{statmorph} morphology
contain only essential columns from the corresponding structures of the candidate catalogue.
The technical and unimportant information was omitted.

For the sake of convenience the information is gathered in two tables computed on the fly: 
a table with general information about objects 
and a table with the \textsc{SExtractor} photometry and the \textsc{statmorph} parameters of galaxies in the five Pan-STARRS bands.
These tables form the catalogue of the nearly edge-on galaxies in Pan-STARRS.

The general information table includes the following data:
\begin{description}
\item[\texttt{pgc}] -- unique identification number linked to the HyperLeda database \citep{2014A&A...570A..13M};
\item[\texttt{egips}] -- EGIPS designation;
\item[\texttt{ra}, \texttt{dec}, \texttt{J2000}, \texttt{fcoo}] 
  -- Right Ascension and Declination in degrees for the J2000.0 epoch, 
  as well as the coordinates in the sexagesimal format; 
  and the coordinates quality flag;
\item[\texttt{E(B-V)}] -- colour excess according to the Galactic extinction map by \citet{1998ApJ...500..525S};
\item[\texttt{votes}, \texttt{pctgood}, \texttt{pctacceptable}, \texttt{pctunsuitable}] 
  -- number of votes during the visual inspection
  and percentage of votes for edge-on (good), nearly edge-on (acceptable), and not edge-on (unsuitable) orientation of galaxies
  (see Section~\ref{sec:CandidateSelection});
\item[\texttt{rfgc}] -- cross-identification with RFGC~\citep{1999BSAO...47....5K};
\item[\texttt{egis}] -- cross-identification with EGIS~\citep{2014ApJ...787...24B};
\item[\texttt{leda}] -- principal object name, \texttt{objname}, in the HyperLeda database \citep{2014A&A...570A..13M};
\item[\texttt{vmax}] -- apparent maximum rotation velocity of the gas in km\,s$^{-1}$ (HyperLeda: \texttt{vmaxg});
\item[\texttt{cz}] -- heliocentric redshift, $cz$, in km\,s$^{-1}$ (HyperLeda: \texttt{v});
\item[\texttt{cz3k}] -- redshift in the CMB rest frame in km\,s$^{-1}$ (HyperLeda: \texttt{v3k});
\end{description}

The photometry table collects the \textsc{SExtractor} data~\citep{1996A&AS..117..393B} 
together with \textsc{statmorph} statistics~\citep{2019MNRAS.483.4140R}:
\begin{description}
\item[\texttt{pgc}] -- Leda identification number;
\item[\texttt{projcell}, \texttt{subcell}] -- projection cell and sub-cell numbers indicating the skycell in the Pan-STARRS sky tessellation;
\item[\texttt{candidate}] -- candidate number inside the given skycell; 
  a triplet of numbers (\texttt{projcell}, \texttt{subcell}, \texttt{candidate}) uniquely identifies an object in the candidate catalogue;

\item[\texttt{band}] -- character field indicating one of the five Pan-STARRS bands: $g$, $r$, $i$, $z$, $y$;

\item[\texttt{ra}, \texttt{dec}] 
  -- Right Ascension and Declination in degrees for the J2000.0 epoch;

\item[\texttt{a}, \texttt{e\_a}] 
  -- standard deviation of the distribution of light along the major axis of the galaxy (\textsc{SExtractor}: \texttt{A}) and its uncertainty in arcsec. For convenience, we call it the semi-major axis, although by definition it is a light distribution scale-length.
\item[\texttt{b}, \texttt{e\_b}] 
  -- standard deviation of the distribution of light along the minor axis of the galaxy (\textsc{SExtractor}: \texttt{B}) and its uncertainty in arcsec. For convenience, we call it the semi-minor axis of the object.
\item[\texttt{ell}] -- ellipticity $ = 1 - b/a$ (\textsc{SExtractor}: \texttt{ELLIPTICITY});
\item[\texttt{pa}] 
  -- position angle of the major axis (\textsc{SExtractor}: \texttt{THETA\_J2000}) 
  measured counterclockwise from the North direction (J2000);

\item[\texttt{radkron}] 
  -- reduced Kron pseudo-radius (\textsc{SExtractor}: \texttt{KRON\_RADIUS});
\item[\texttt{magauto}, \texttt{e\_magauto}] 
  -- estimate of the total apparent magnitude (\textsc{SExtractor}: \texttt{MAG\_AUTO}) 
  and its error using the Kron's `first moment' algorithm \citep{1980ApJS...43..305K};
\item[\texttt{magautocor}] 
  -- the Kron magnitude corrected for Galactic extinction \citep{2011ApJ...737..103S};

\item[\texttt{radpetro}] 
  -- reduced Petrosian pseudo-radius (\textsc{SExtractor}: \texttt{PETRO\_RADIUS});
\item[\texttt{magpetro}, \texttt{e\_magpetro}] 
  -- Petrosian total apparent magnitude (\textsc{SExtractor}: \texttt{MAG\_PETRO}) 
  and its error \citep{1976ApJ...209L...1P};
\item[\texttt{magpetrocor}] 
  -- the Petrosian magnitude corrected for Galactic extinction \citep{2011ApJ...737..103S};
  
\item[\texttt{badpixfraction}] -- fraction of bad pixels inside the Petrosian ellipse;
\item[\texttt{quality}] -- set to `false' if there are indications of photometry problems;

\item[\texttt{gini}] 
  -- the Gini coefficient \citep{2004AJ....128..163L} measures the inequality of the pixel flux value distribution over a galaxy
  \citep[for details see][]{2019MNRAS.483.4140R}.
  A Gini coefficient of zero means perfect equality (all pixels of a galaxy have the same flux),
  while a Gini coefficient of one indicates total inequality (the whole flux is concentrated in just one pixel).
\item[\texttt{m20}] 
  -- the $M_{20}=\log \mu_{20}/\mu_{\rm tot}$ statistics \citep{2004AJ....128..163L} 
  is the second-order moment $\mu=\sum I_ir_i^2$ of the brightest 20 per cent of the total flux 
  normalized to the total second-order central momentof the galaxy, 
  where $I_i$ is the flux at the $i^{\rm th}$ pixel and $r_i$ is its distance from the galaxy center
  \citep[for details see][]{2019MNRAS.483.4140R}.
  It tracks bright structures in the galaxy:
  compact nuclei have small $M_{20}$ values, while bars and spiral arms produce high $M_{20}$.
\item[\texttt{concentration}] 
  -- the concentration index, $C=5\log r_{80}/r_{20}$,
  where $r_{20}$ and $r_{80}$ are the radii of circular apertures 
  containing 20 and 80 per cent of the total light of the object
  \citep[for details see][]{2019MNRAS.483.4140R}.
\item[\texttt{asymmetry}]
  -- the asymmetry index is calculated by subtracting a 180 degree rotated galaxy image from itself 
  \citep[for details see][]{2019MNRAS.483.4140R}.
\item[\texttt{smoothness}]
  -- smoothness index (also known as `clumpiness') is the difference between the original and the smoothed image of the galaxy
  \citep[for details see][]{2019MNRAS.483.4140R}.

\end{description}

\section{Completeness}

\begin{figure}
\centering
\includegraphics[width=0.45\textwidth]{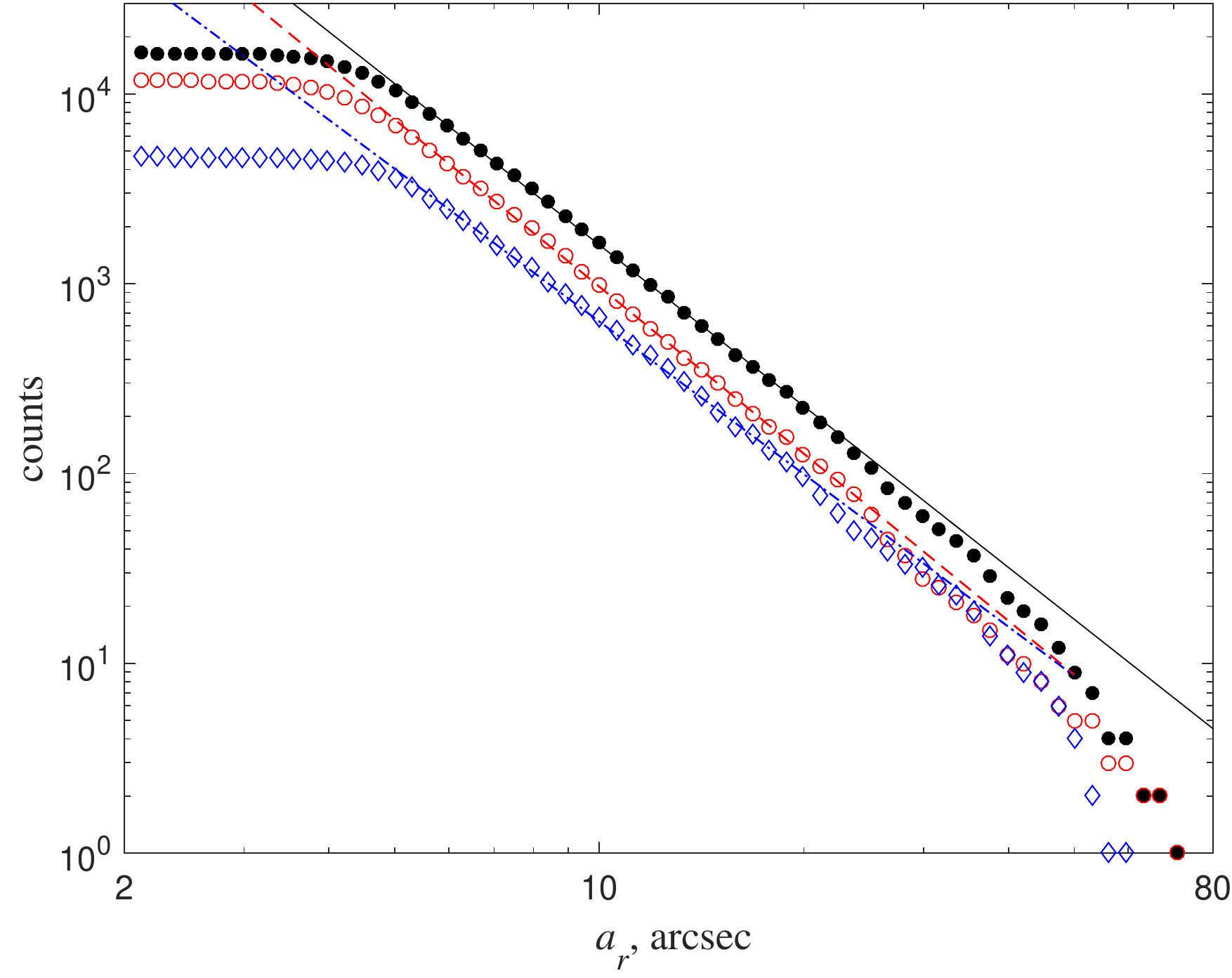}
\caption{
Completeness function $\log N$ -- $\log a_r$ for EGIPS galaxies.
Black dots represent the distribution for the whole sample.
Red open circles correspond to the red galaxies $(g-i)_0>0.95$, 
while blue diamonds illustrate the behaviour of the blue galaxies.
}
\label{fig:logNlogA}
\end{figure}

To estimate the completeness of the edge-on galaxy catalogue, we used two simple tests.
Using the major axis scale-length as a substitute for the galaxy diameter, 
in Fig.~\ref{fig:logNlogA} we plotted the cumulative number of galaxies 
as a function of their angular sizes in the $r$ band.
In the case of a uniform distribution, the slope of the $\log N$--$\log a_r$ relation
should be equal to $-3$.
The linear part of this completeness function for the entire sample of edge-on galaxies
follows the relation $\log N \propto (-2.82\pm0.02) \log a_r$.
The population of the red galaxies, $(g-i)_0>0.95$, demonstrates the slope of $-2.92\pm0.02$,
and the blue galaxies show a slope of $-2.67\pm0.02$.
The graphics allow us to estimate the completeness of the catalogue for the objects with $a_r>5.5$~arsec 
at the level of 96 per cent.
Also we performed the $V/V_m$ test, originally developed by \citet{1979ApJ...231..680T}.
The test shows that the sample is essentially complete, $V/V_m = 0.461$, 0.472 and 0.476,
for the objects with $a>6$~arcsec in the $g$, $r$ and $i$ bands, respectively.

\section{Sky and redshift distribution}

\begin{figure}
\centering
\includegraphics[width=0.45\textwidth,bb=97 66 678 357,clip]{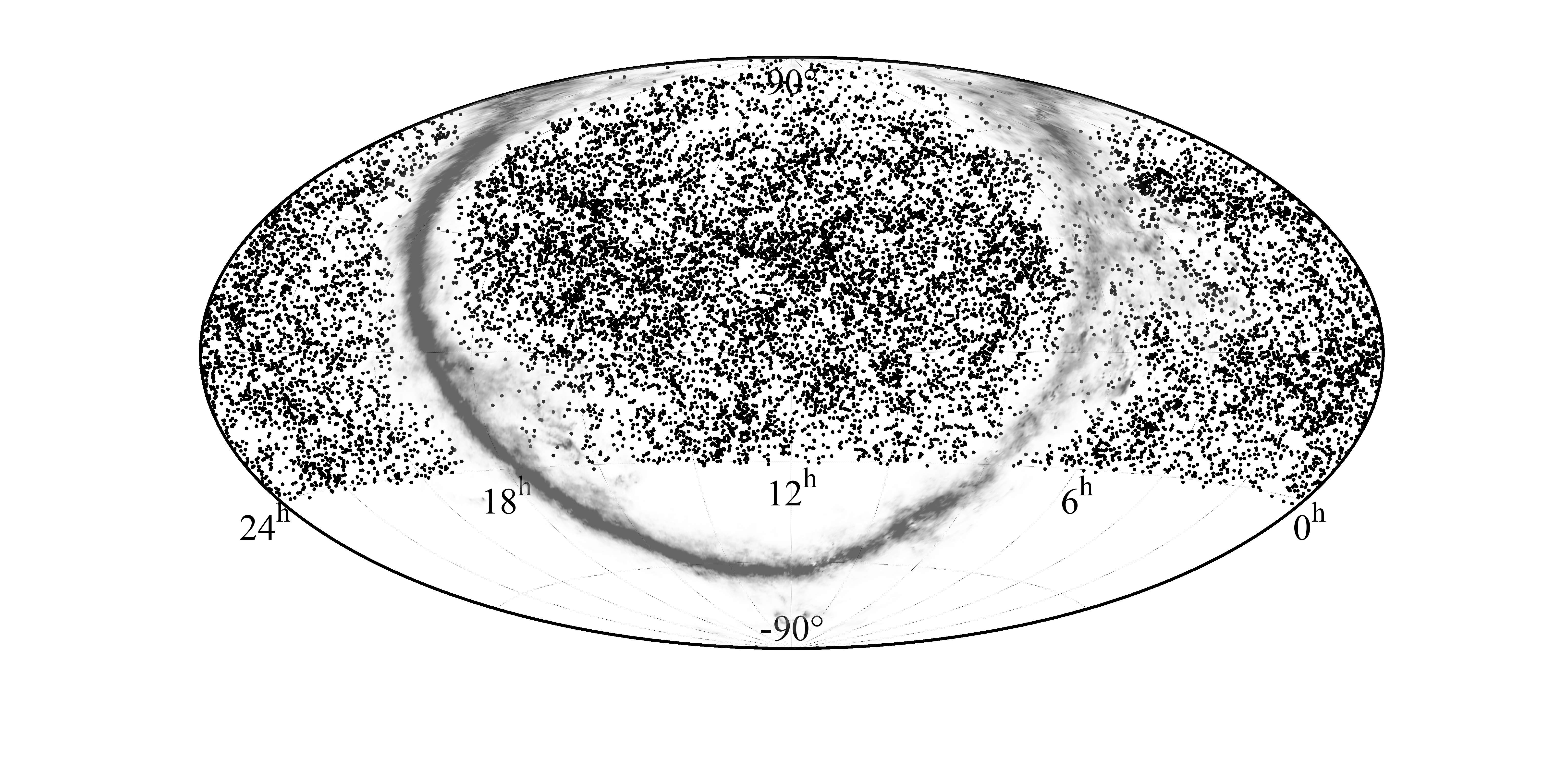}
\caption{
Distribution of edge-on galaxies from our catalogue over the sky in the equatorial coordinate system.
The fuzzy grey belt represents an extinction map for our Galaxy.
}
\label{fig:skymap}
\end{figure}

Figure~\ref{fig:skymap} shows the sky distribution of our edge-on galaxies in the equatorial coordinate system.
The Pan-STARRS survey is limited to $\textrm{Dec.}>-30\degr$ which is reflected on our map.
The main feature is the area of strong Galactic extinction shown in gray where the number of galaxies drops to almost zero.
As can bee seen, the sky distribution of edge-on galaxies is nearly random.
The plane of the Local supercluster is barely seen.
The region of the Local void is well filled with galaxies, 
which means that the depth of our catalogue is significantly deeper 
than the size of the Local void of 35--70~Mpc \citep{2008ApJ...676..184T}.

\begin{figure}
\centering
\includegraphics[width=0.48\textwidth]{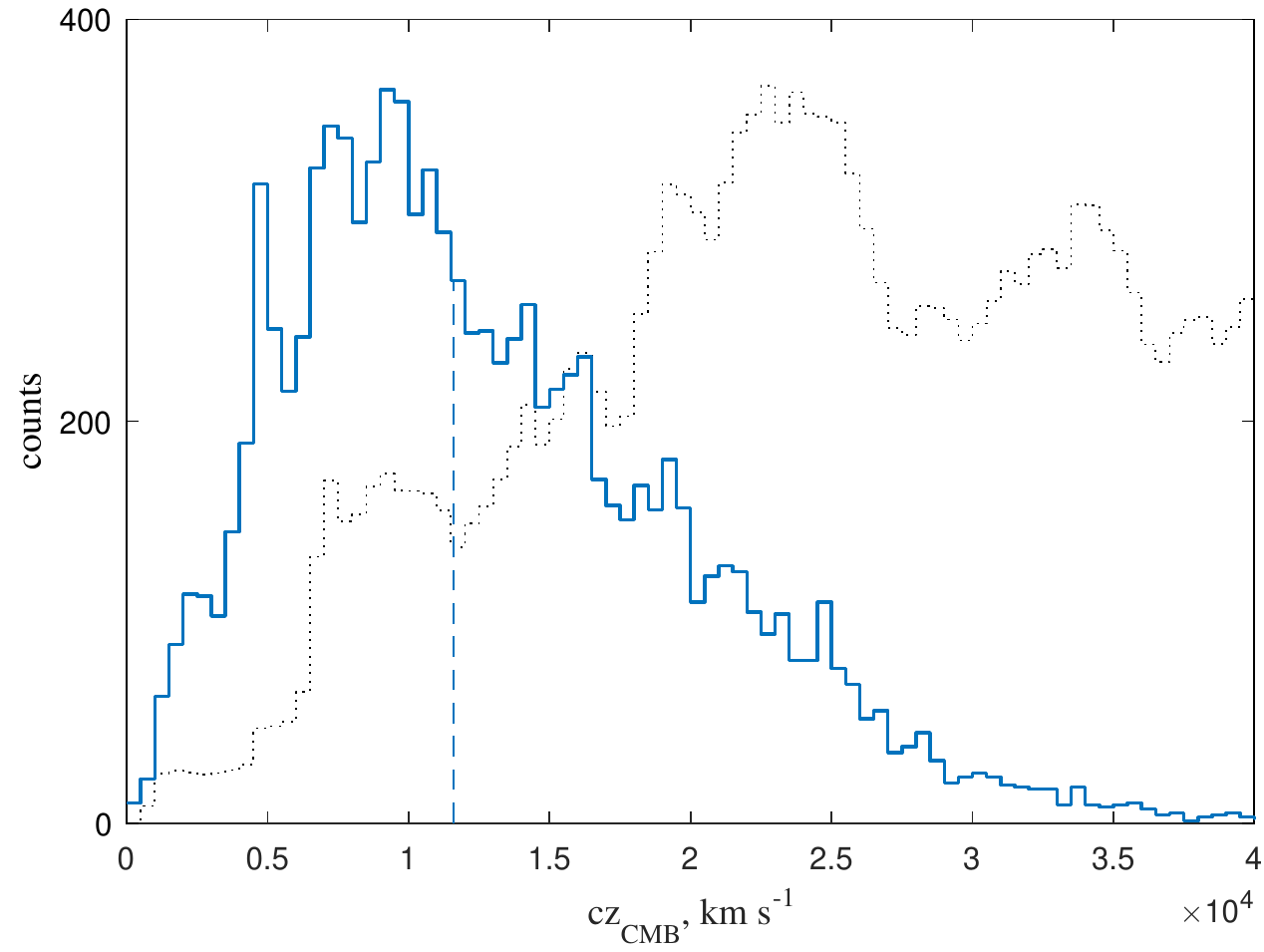}
\caption{
The redshift distribution of edge-on galaxies.
The median value 11,600~km\,s$^{-1}$ is indicated by the vertical dashed line.
For comparison, the dotted line shows the distribution of SDSS galaxies scaled to fit the y-axis.
}
\label{fig:cz}
\end{figure}

According to HyperLeda, 10,485 of our 16,551 galaxies already have measured redshifts.
Taking into account only known data, the effective depth of the catalogue is characterized 
by the median velocity of 11,600~km\,s$^{-1}$ in the CMB frame of reference,
which roughly corresponds to 165--170~Mpc.
According to Fig.~\ref{fig:cz},
the redshift distribution reveals a strong shortage of galaxies above $cz_{\rm CMB}\sim10,000$~km\,s$^{-1}$.

\begin{figure}
\centering
\includegraphics[width=0.45\textwidth]{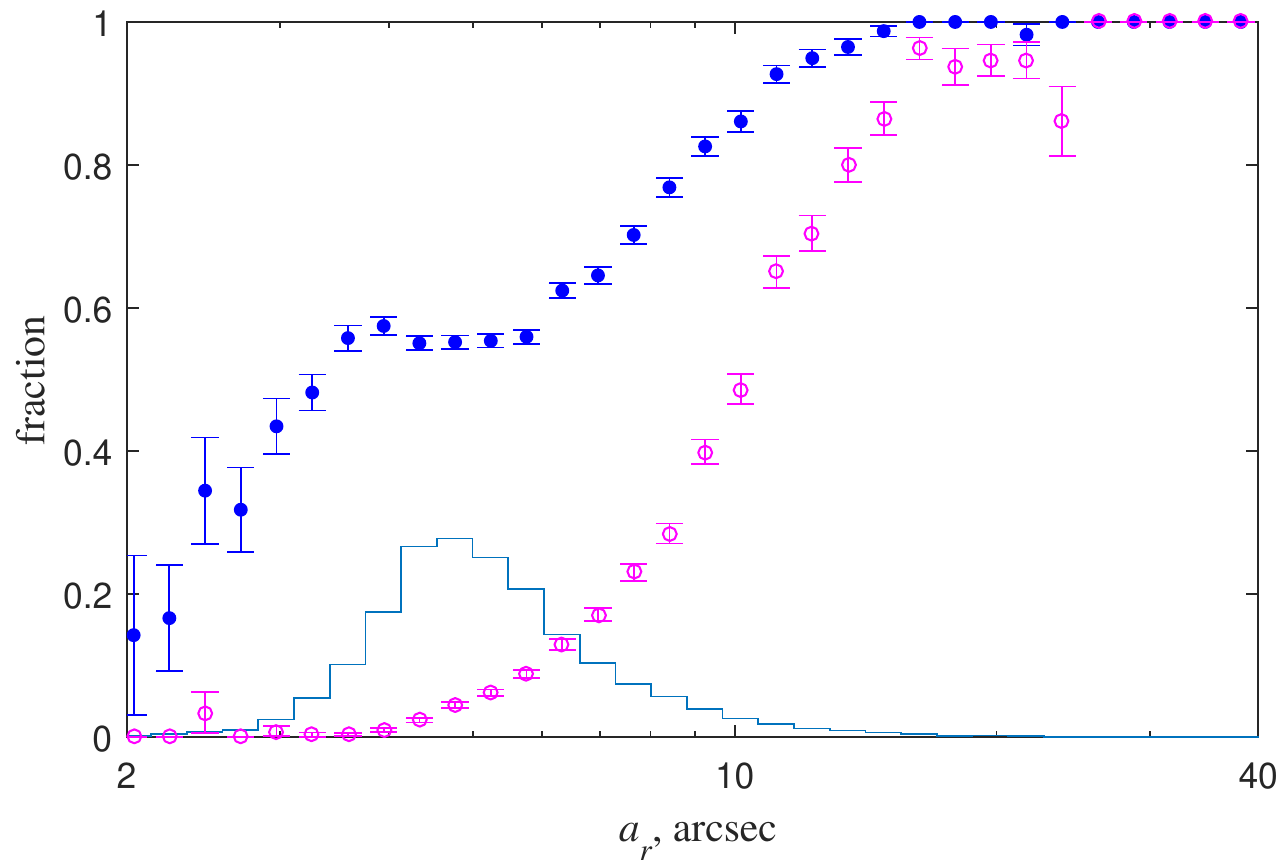}
\includegraphics[width=0.45\textwidth]{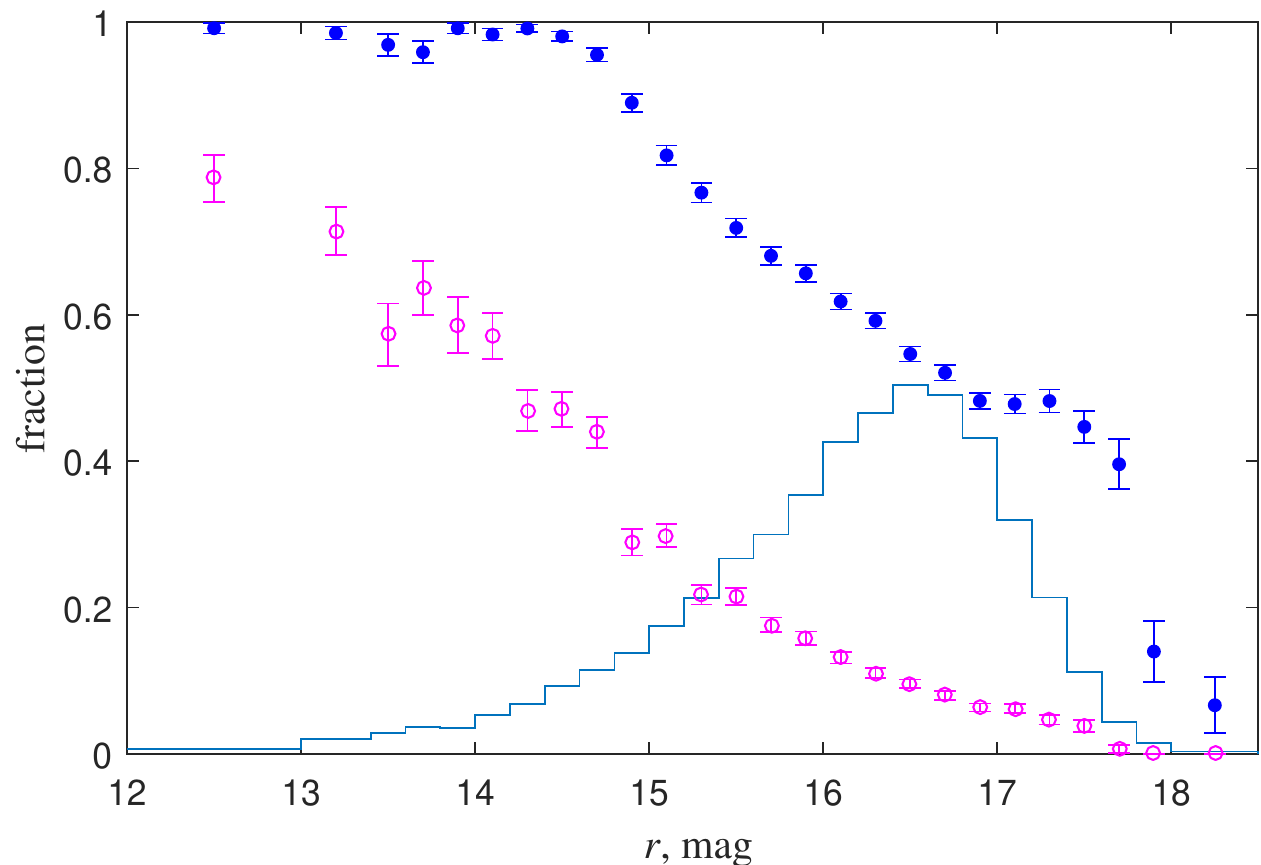}
\includegraphics[width=0.45\textwidth]{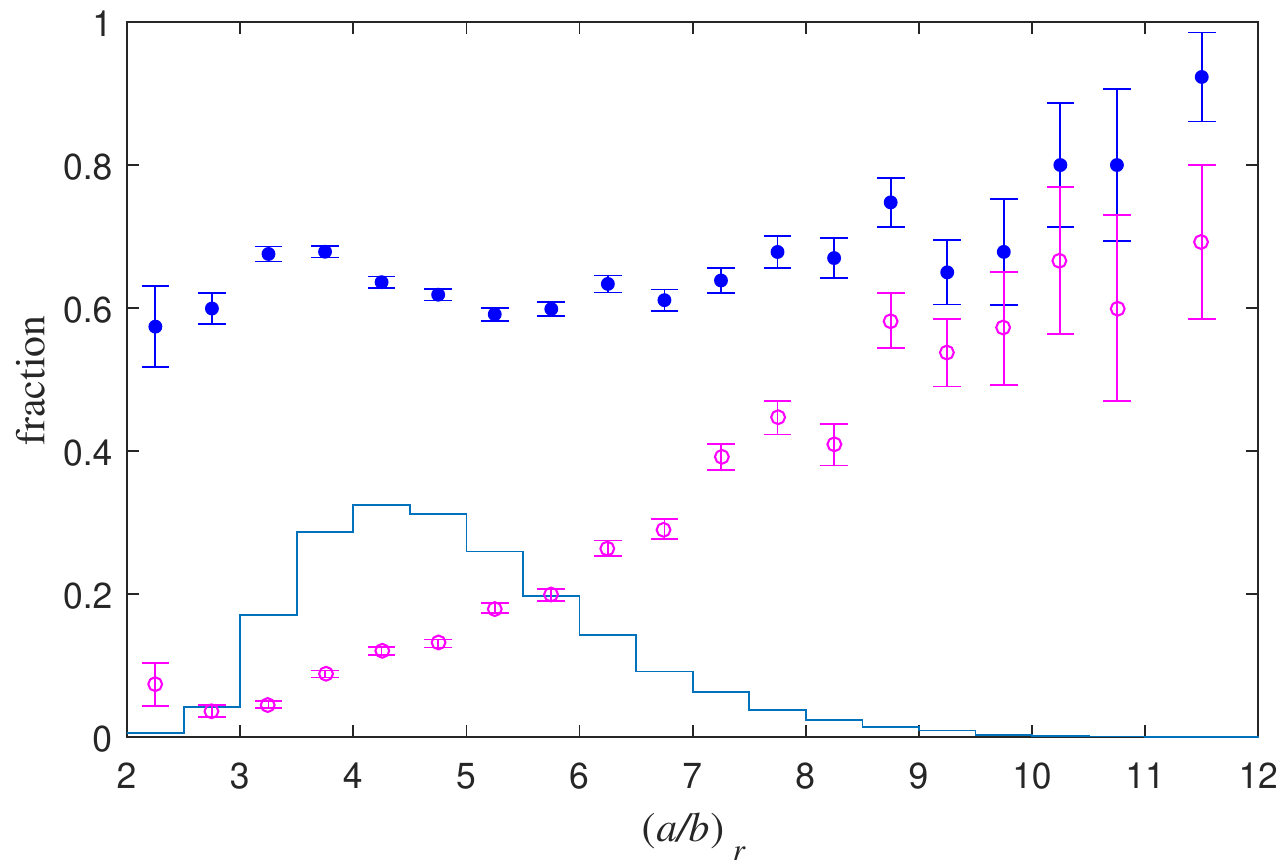}
\caption{
Fraction of edge-on galaxies with known redshifts and internal kinematics.
Each panel shows the completeness as a function of the redshift (blue error bars) 
and internal kinematics (magenta error bars with open circles) 
for edge-on galaxies as function of the major semi-axis (top panel), 
total Petrosian magnitude (middle panel), and the axes ratio (bottom panel) in the $r$ band.
The probability distribution of galaxies is shown by the histogram.
}
\label{fig:czCompleteness}
\end{figure}

As expected, the redshift data are 100 per cent complete for the large, $a_r\gtrsim14$~arcsec, 
and bright, $r_{\rm Petro}\lesssim14.6$~mag, galaxies (see Fig.~\ref{fig:czCompleteness}).
The completeness gradually decreases with decreasing galaxy size and reaches the level of 50 per cent 
for galaxies with $a_r\approx3.3$~arcsec and $r_{\rm Petro}\approx17.3$~mag.

The situation with the internal kinematic data is much poorer.
The radio 21~cm line widths and optical rotation curves are available only for 2800 galaxies, 
17 per cent of the sample.
The fraction of galaxies with available internal kinematics 
is shown in Fig.~\ref{fig:czCompleteness} by the magenta error bars.
The data are almost complete only for most extended objects, $a_r\gtrsim15.5$~arcsec (the top panel).
The internal kinematics completeness drops rapidly 
and for galaxies $a_r\lesssim10.5$~arcsec turns out to be less than 50 per cent.
Since performing spectroscopic observations of the rotation of early-type galaxies is much more difficult 
than measuring the \ion{H}{I} line width in gas-rich galaxies, 
as a result, much more data are available on the maximum rotational velocity of late-type spiral galaxies.
This explains the relatively small percentage of rotation curves available 
even for the brightest galaxies (the middle panel)
and the continuous growth of the completeness towards the thinnest galaxies (the bottom panel).

\section{Comparison with the EGIS and RFGC catalogues}
\label{sec:Comparison}

\begin{figure}
\centering
\includegraphics[width=0.48\textwidth]{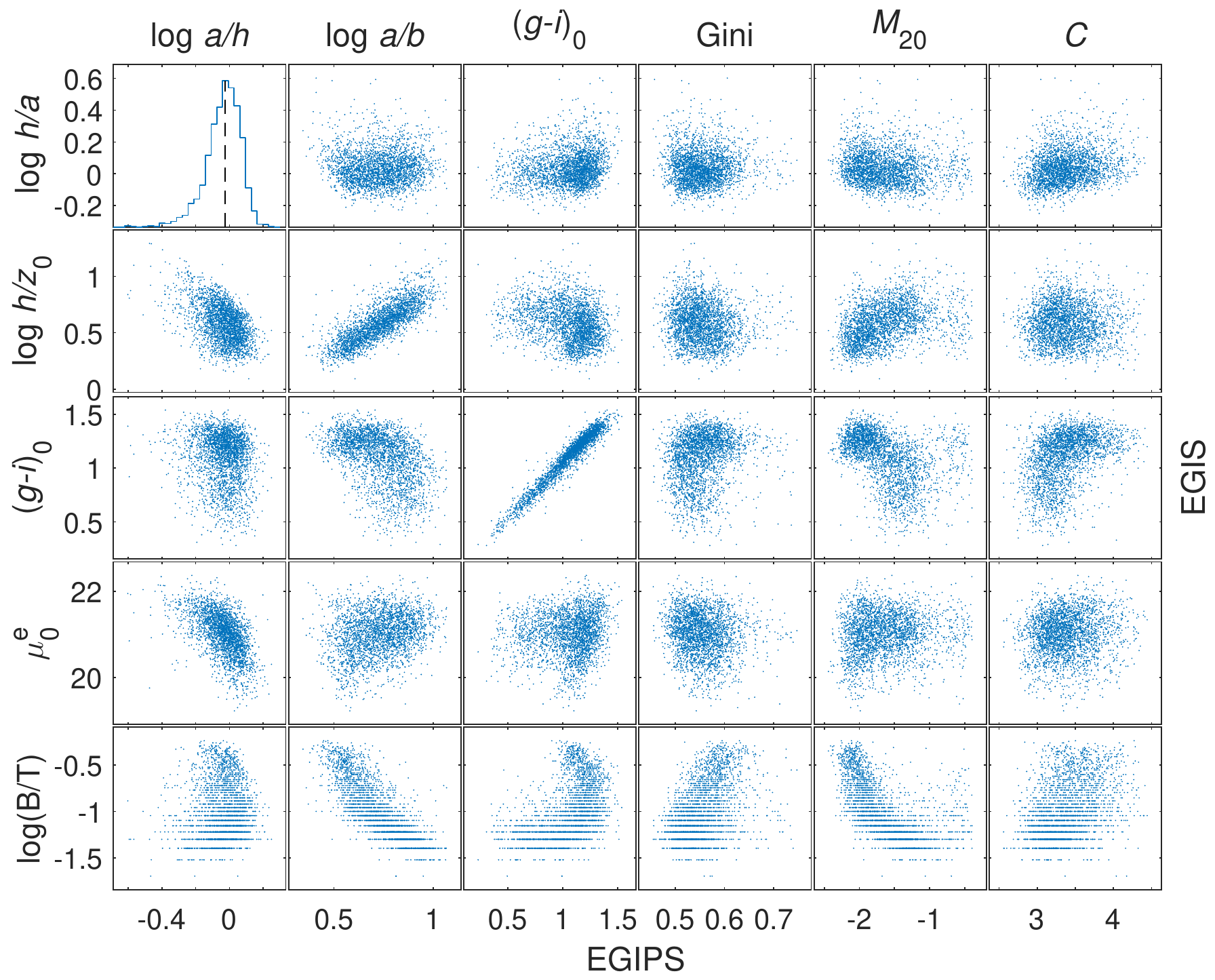}
\caption{
Comparison of the properties for cross-correlated galaxies in EGIS and EGIPS in the $g$ band.
Our Pan-STARRS photometry is presented horizontally.
The photometric parameters from the EGIS aperture photometry are arranged vertically.
}
\label{fig:correlations}
\end{figure}

\begin{table*}
\centering
\caption{
The size relations between the galaxy disc model from EGIS and the \textsc{SExtrator} photometry.
The first two blocks give the transformation from disc parameters in the EGIS catalogue 
to the size estimates derived in the current work.
The last two blocks give the inverse transformation from our photometry 
to the EGIS disc scale-lengths.
The response variable and corresponding regression are indicated in the first row of each table block.
The first column specifies the passband.
The second column gives the median of the response variable values.
The last column contains the estimate of the residual scatter.
The other columns show the coefficients of a robust multilinear regression for the specified predictors
(only statistically significant coefficients are shown).
}
{
\footnotesize
\begin{tabular}{ccccccc}
\hline\hline
band& Median   & $k_0$            & $k_1$              & $k_2$              & $k_3$            & $\sigma$ \\
\hline
\multicolumn{7}{l}{$\log a - \log h = k_0 + k_1 \log h/z_0 + k_2 \mu_0^e + k_3 \log B/T$}\\[2pt]
$g$ & $-0.026$ & $+2.003\pm0.050$ & $-0.4333\pm0.0095$ & $-0.0948\pm0.0024$ & $-0.2098\pm0.0052$ & 0.059 \\
$r$ & $+0.029$ & $+1.410\pm0.038$ & $-0.3929\pm0.0078$ & $-0.0665\pm0.0020$ & $-0.1943\pm0.0041$ & 0.047 \\
$i$ & $+0.045$ & $+1.262\pm0.033$ & $-0.3659\pm0.0074$ & $-0.0596\pm0.0017$ & $-0.1836\pm0.0038$ & 0.045 \\

\hline
\multicolumn{5}{l}{$\log a/b - \log h/z_0 = k_0 + k_1 \log h/z_0 + k_2 (g-r)_0 + k_3\log B/T$} \\[2pt]
$g$ & $+0.155$ & $+0.3831\pm0.0093$ & $-0.6257\pm0.0094$ & $-0.0841\pm0.0084$ & $-0.1883\pm0.0058$ & 0.059 \\
$r$ & $+0.155$ & $+0.3681\pm0.0076$ & $-0.5954\pm0.0081$ & $-0.0502\pm0.0072$ & $-0.1638\pm0.0049$ & 0.049 \\
$i$ & $+0.157$ & $+0.3725\pm0.0075$ & $-0.5890\pm0.0081$ & $-0.0481\pm0.0072$ & $-0.1551\pm0.0049$ & 0.048 \\

\hline
\multicolumn{5}{l}{$\log h-\log a = k_0 + k_1 \log a/b + k_2 \mathrm{Gini} + k_3 C $} \\[2pt]
$g$ & $+0.026$ & $-0.104\pm0.025$ &                  & $-0.331\pm0.049$ & $+0.0935\pm0.0066$ & 0.094 \\
$r$ & $-0.029$ & $-0.267\pm0.019$ & $+0.061\pm0.012$ &                  & $+0.0577\pm0.0046$ & 0.074 \\
$i$ & $-0.045$ & $-0.315\pm0.021$ &                  & $+0.212\pm0.038$ & $+0.0421\pm0.0051$ & 0.069 \\

\hline
\multicolumn{5}{l}{$\log h/z_0-\log a/b = k_0 + k_1 \log a/b + k_2 (r-i)_0 + k_3 C $} \\[2pt]
$g$ & $-0.155$ & $-0.369\pm0.024$ & $+0.069\pm0.014$ & $+0.163\pm0.022$ & $+0.0315\pm0.0060$ & 0.093 \\
$r$ & $-0.155$ & $-0.393\pm0.021$ & $+0.142\pm0.014$ & $+0.196\pm0.019$ & $+0.0197\pm0.0053$ & 0.082 \\
$i$ & $-0.157$ & $-0.350\pm0.013$ & $+0.145\pm0.014$ & $+0.255\pm0.017$ &                    & 0.080 \\
\hline\hline
\end{tabular}
}
\label{tab:SizeRelations}
\end{table*}

The photometric parameters in our catalogue are obtained from an analysis of the overall light distribution in a galaxy.
For instance, the major and minor semi-axes, $a$ and $b$, 
are close in meaning to the scale-lengths of the light distribution, 
but for the entire galaxy, and not for its individual structural components.
Fortunately, our sample of edge-on galaxies contains a significant number of objects 
from the RFGC (2237) and EGIS (3231) catalogues,
which allows us to compare our parameters with those measured in the literature.
Note that about 10 per cent of RFGC (242 of 2479) and EGIS (366 of 3597) galaxies did not pass our visual inspection
and were not included into the final sample of edge-on galaxies.
The structural parameters of the galactic discs 
(the radial scale-length, the vertical scale-height, the central surface brightness etc.) 
in the EGIS catalogue were derived from the bulge-disc decomposition 
of the surface brightness profiles based on SDSS DR7 images \citep{2014ApJ...787...24B}.
The RFGC catalogue \citep{1999BSAO...47....5K} contains only measurements of the isophotal diameters. 
However, due to the selection criteria $a/b>7$, it is populated mainly by late-type bulgeless galaxies.
Thus, the RFGC sizes reflect the isophotal disc diameters quite reasonably.

In Fig.~\ref{fig:correlations} 
we compare distance-independent properties in the $g$ band obtained in this article with those in EGIS.
The first column and the first row show the ratio of two ways to measure the galaxy sizes, 
the $\log h/a$ ratio, and the complementary value $\log a/h$, respectively.
Here $h$ is the exponential disc scale-length from the profile fitting in EGIS
and $a$ is the standard deviation of the light distribution along the major axis from the \textsc{SExtractor} photometry.
The other columns contain the data from the EGIPS catalogue (this work):
$\log a/b$ is the major to minor axis ratio;
$(g-i)_0$ is the total galaxy colour from the Petrosian magnitudes 
corrected for Galactic extinction~\citep{2011ApJ...737..103S};
Gini, $M_{20}$ and concentration ($C$) are the non-parametric morphological indexes described in Section~\ref{sec:Catalogue}.
The rows correspond to the data from the EGIS:
$\log h/z_0$ is the inverse of the thickness of the disc, 
where $z_0$ is the vertical scale-height of the $\mathrm{sech}^2$ disc; 
$(g-i)_0$ is the total galaxy colour from the aperture photometry corrected for Galactic reddening;
$\mu_0^e$ is the central surface brightness of the edge-on disc;
$\log B/T$ is the bulge-to-total light ratio.

The top left panel of Fig.~\ref{fig:correlations} shows the concordance of the radial scale-lengths 
of the EGIS and \textsc{SExtractor} photometry.
The median value of $\log h/a = 0.026$ in $g$ band means 
that the \textsc{SExtractor} estimate of the standard deviation of the light distribution along the major axis, $a$, 
gives a very good proxy of the exponential disc scale-length, $h=1.06a$ with a spread of 0.094~dex.
The concordance is retained in other bands as well: 
$\log h/a = -0.029$ with a spread of 0.072
and $\log h/a = -0.045$ with a spread of 0.068~dex in the $r$ and $i$ bands, respectively
(see the second column, {\it Median}, in Table~\ref{tab:SizeRelations}).
The top row shows that this ratio depends on the morphology and colour of the galaxy.
$\log h/a$ correlates most strongly with the concentration index.
It is a good surprise, because the concentration index is determined in circular apertures 
and there was some concern about whether it is suitable for highly flattened objects as edge-on galaxies.

The axis ratio $\log a/b$ (the second column of Fig.~\ref{fig:correlations}), 
measured with \textsc{SExtractor}, 
is systematically higher than the disc scale-length-to-height ratio (the second row) from the model fitting, 
the median of $\log a/b - \log h/z_0 = 0.155$.
This value is quite stable in all passbands under consideration.

The tightest correlation is found for the colours (the third row and the third column of Fig.~\ref{fig:correlations}).
It reflects the fact of surprisingly good photometry performed by \textsc{SExtractor}  
for the Pan-STARRS images (see Section~\ref{sec:Photometry}),
despite known problems with processing extended and elongated objects.

It is noteworthy that the colour-morphology indexes and the colour-thickness diagrams clearly show 
the segregation of galaxies into different populations.
The well-known separation of galaxies in the colour-magnitude diagram is discussed in Section.~\ref{sec:GalaxyCMD}.

The relations between the parameters in EGIS and EGIPS are summarised in Table~\ref{tab:SizeRelations}.
The coefficients were estimated using a robust multilinear regression.
Note that the parameters from the model fitting give a significantly better prediction
of the parameters obtained by \textsc{SExtractor} than the usage of the non-parametric morphology in the opposite relations.
Therefore, an accurate knowledge about the bulge--disc decomposition significantly improves the relationship, 
namely the prediction of $a$.
Using the disc thickness, $\log h/z_0$, the edge-on disc central surface brightness, $\mu_0^e$, 
and the bulge-to-total light ratio, $\log B/T$, (the first row) 
reduces the scatter to 0.059, 0.047 and 0.045~dex in the $g$, $r$ and $i$ bands, respectively.
Here we use the central surface brightness for an edge-on orientation, as seen from observations,
because the correction $+2.5\log h/z_0$ to the face-on central surface brightness induces an artificial correlation 
with the galaxy thickness, as noted by \citet{2014MNRAS.441.1066M}. 

A similar comparison allows us to find a connection between our \textsc{SExtractor}'s ellipses derived in Pan-STARRS images
and the isophotal diameters of the flat RFGC galaxies measured on photographic films of POSS-I and ESO/SERC surveys.
The medians of the typical RFGC semi-axes are $a_O=3.43\,a_g$ and $a_E = 3.09\,a_r = 3.12\,a_i$.
The medians of the axis ratios are $(a/b)_O=1.29\,(a/b)_g$ and $(a/b)_E = 1.21\,(a/b)_r = 1.24\,(a/b)_i$.
Here, the subscripts `$O$' and `$E$' denote the blue and red photographic plates used in the RFGC,
while the subscripts `$g$', `$r$' and `$i$' refer to the Pan-STARRS bands.

\section{Galaxy colour-magnitude diagram}
\label{sec:GalaxyCMD}

\begin{figure}
\centering
\includegraphics[width=0.45\textwidth]{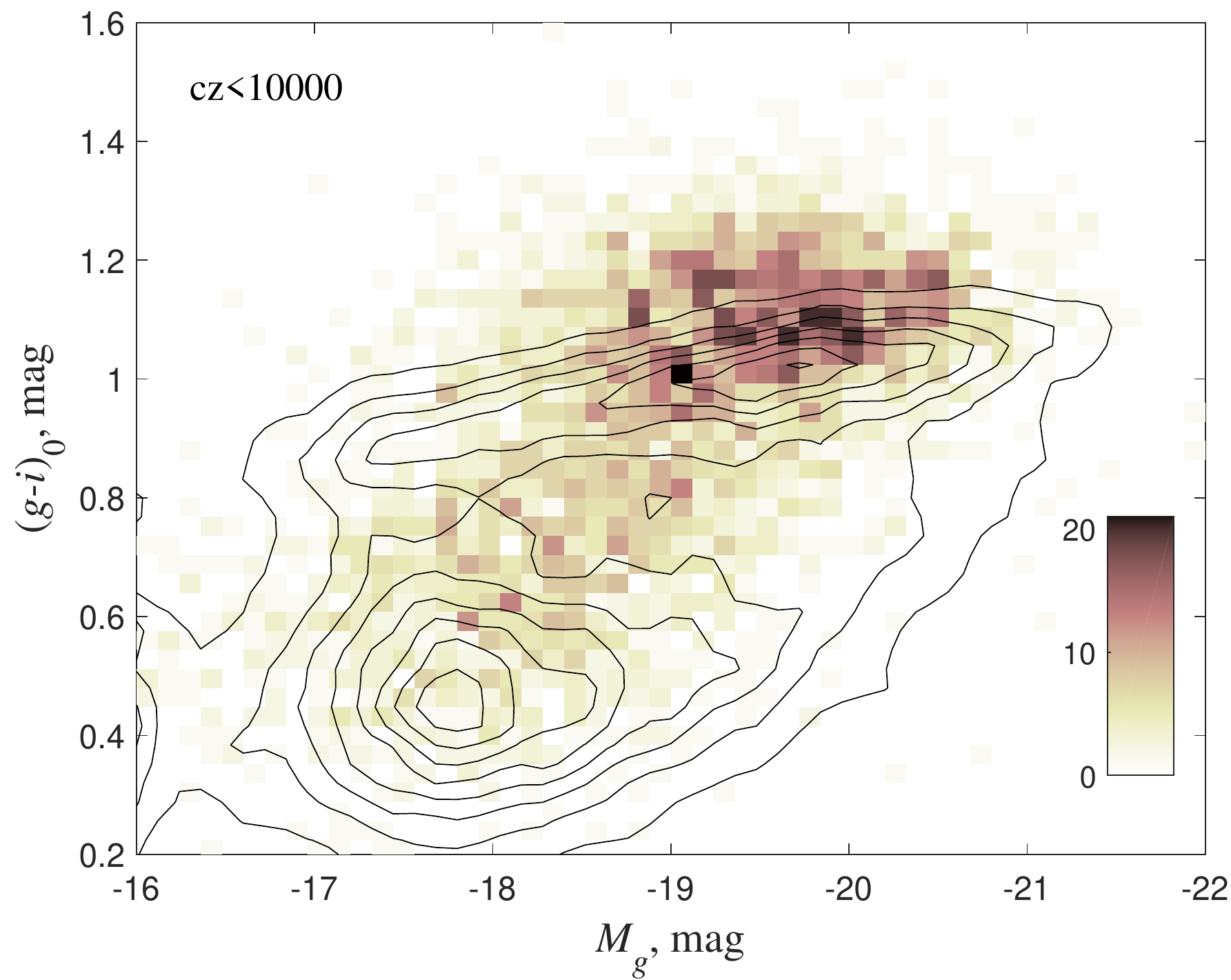}
\includegraphics[width=0.45\textwidth]{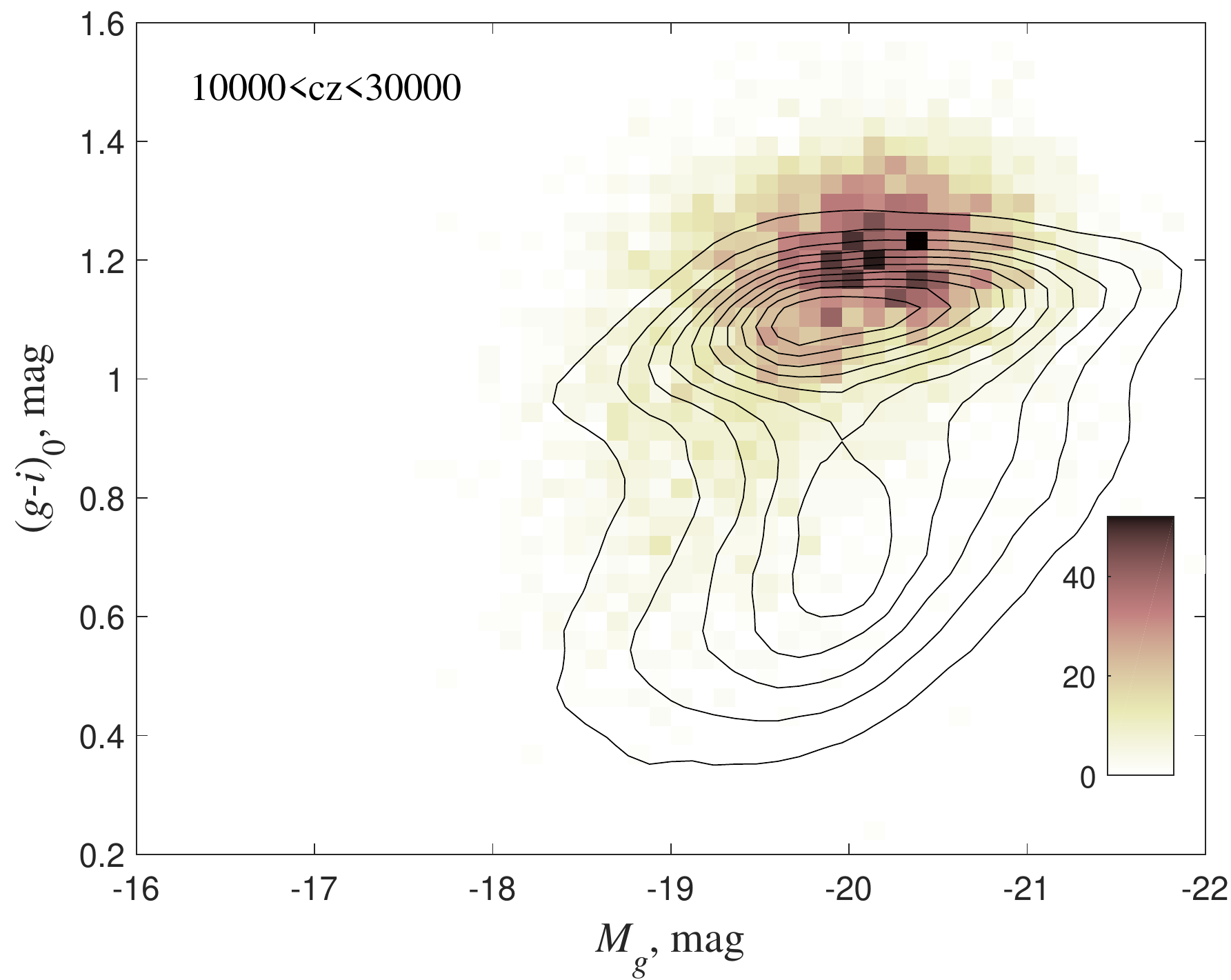}
\caption{
The galaxy colour--magnitude diagram.
The density distribution of the EGIPS galaxies is shown in brown.
The solid isolines illustrate the distribution of a general sample of nearby galaxies,
taken from the SDSS survey and transformed to the Pan-STARRS photometric system.
All magnitudes have been corrected for Galactic extinction.
}
\label{fig:galaxyCMD}
\end{figure}

A general population of galaxies shows a bimodal distribution in the colour--absolute magnitude diagram.
\citet{2004ApJ...608..752B} were the first to describe the individual areas of the diagram.
The red sequence is populated by early-type galaxies while 
the blue cloud is formed by spirals.
There is a transitional area between them, the so-called green valley,
which is interpreted as a zone of fast evolution from late- to early-type galaxies.
The colour--magnitude diagram for edge-on galaxies is shown in Fig.~\ref{fig:galaxyCMD}.
The brown colour represents the density distribution of the EGIPS galaxies.
For comparison, we superimposed a general distribution of galaxies with known redshifts,
taken from the SDSS~DR12~\citep{2015ApJS..219...12A}.
The sample was split into the nearby $cz_{\rm CMB}<10,000$ (top panel) 
and the more distant $10,000<cz_{\rm CMB}<30,000$~km\,s$^{-1}$ (bottom panel) subsamples.
The most nearby galaxies with $cz_{\rm CMB}<2000$~km\,s$^{-1}$ were excluded from the consideration
to minimize the influence of the local peculiar velocity field 
and to avoid the possible problems of applying automatic photometry to very extended objects.
The magnitudes are corrected for the Galactic extinction~\citep{2011ApJ...737..103S}, 
but are not corrected for the internal extinction.
SDSS Petrosian magnitudes were transformed to the Pan-STARRS photometric system~\citep[][table~6]{2012ApJ...750...99T}.

Fig.~\ref{fig:galaxyCMD} shows that the catalogue is dominated by the red sequence galaxies 
with a colour $(g-i)_0\gtrsim1.0$, which are likely to be lenticular galaxies.
This can partly be explained by a selection effect. 
Unlike spirals, S0-galaxies do not have pronounced features 
that allow for reliable identification of their edge-on orientation.
As a result, this makes the criterion for their selection softer than for the late-type galaxies.

Despite the good agreement between the EGIPS and the general sample of the nearby galaxies, $cz_{\rm CMB}<10,000$, 
there is a noticeable shift in the colour of the red sequence galaxies (top panel of Fig.~\ref{fig:galaxyCMD}).
Edge-on galaxies are $\Delta(g-i)\approx0.1$~mag redder than a typical non-edge-on galaxy.
This can be partially explained by the internal extinction in the edge-on galaxies.
According to \citet{2010MNRAS.404..792M}, the colour difference between the face-on and edge-on orientation is
$(g-i)_0=0.28$ for `Pure disc' and $(g-i)_0=0.20$ for `Very bulgy' galaxies.
Taking into account the random orientation of SDSS galaxies and a typical intrinsic axial ratio of $q\sim0.22$ \citep{2008ApJ...687..976U},
we can estimate the expected colour shift between the general sample and edge-on galaxies to be equal to
$(g-i)_0=0.24$ and 0.16~mag for pure disc and bulge-dominated galaxies, respectively.
However, we should note that our estimation gives less reddening for red sequence edge-on galaxies 
than the value of \citet{2010MNRAS.404..792M}.
There is a rough agreement between the EGIPS and the SDSS galaxies for the blue cloud galaxies, 
but our data do not allow us to reliably measure the effect.

The bottom panel of Fig.~\ref{fig:galaxyCMD} illustrates the observational biases for the distant sample of our galaxies.
As it can be seen from Fig.~\ref{fig:cz}, the redshift distributions of the SDSS and EGIPS galaxies 
differ significantly beyond $cz\sim11,000$~km\,s$^{-1}$.
The number of edge-on galaxies with known redshift drops dramatically, 
while the number of SDSS redshifts continues to grow up to 23,000~km\,s$^{-1}$.
As a result, the subsample of distant $cz_{\rm CMD}>10$,000 general galaxies is dominated 
by high luminosity objects, hence explaining the difference between the EGIPS and SDSS galaxy colour distributions.
This effect is especially prominent for the blue galaxies.

\section{Axis ratio distribution}

\begin{figure}
\centering
\includegraphics[width=0.45\textwidth]{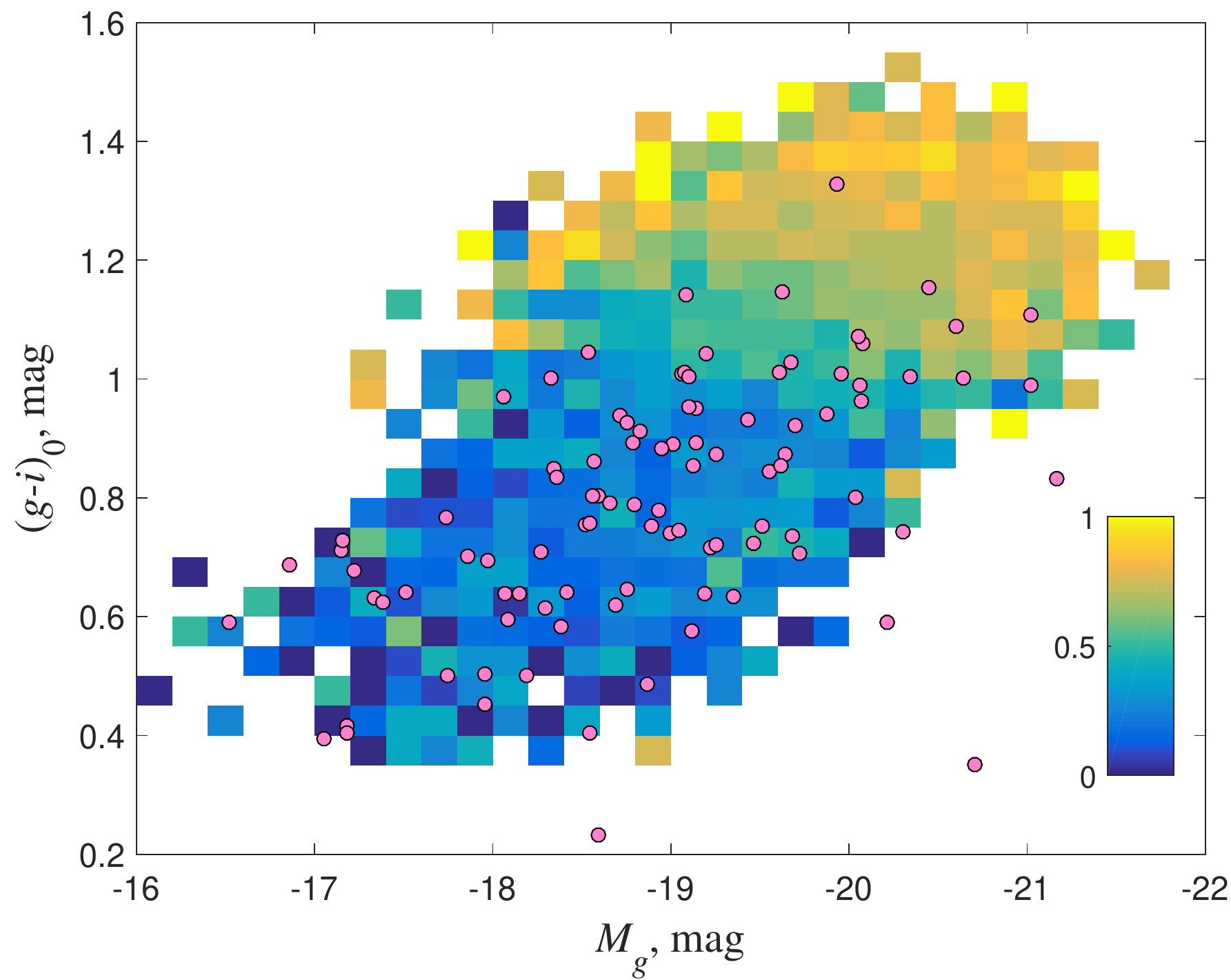}
\caption{
The fraction of thick galaxies, $(a/b)_g<5$, 
over the galaxy colour--magnitude diagram.
The fraction is colour-coded according to the legend.
Only bins with more than 3 galaxies are shown.
The pink dots represent the distribution of superthin galaxies, $(a/b)_g>10$.
}
\label{fig:galaxyCMD+axisRatio}
\end{figure}

The distribution by the axis ratio of edge-on galaxies reveals 
a clear dichotomy in the colour-magnitude diagram shown in Fig.~\ref{fig:galaxyCMD+axisRatio}.
The red sequence is populated by thick, $(a/b)_g<5$, galaxies.
The blue cloud is dominated by thin galaxies.
This segregation probably reflects the morphological differences between these two groups of galaxies.

\begin{figure}
\centering
\includegraphics[width=0.45\textwidth]{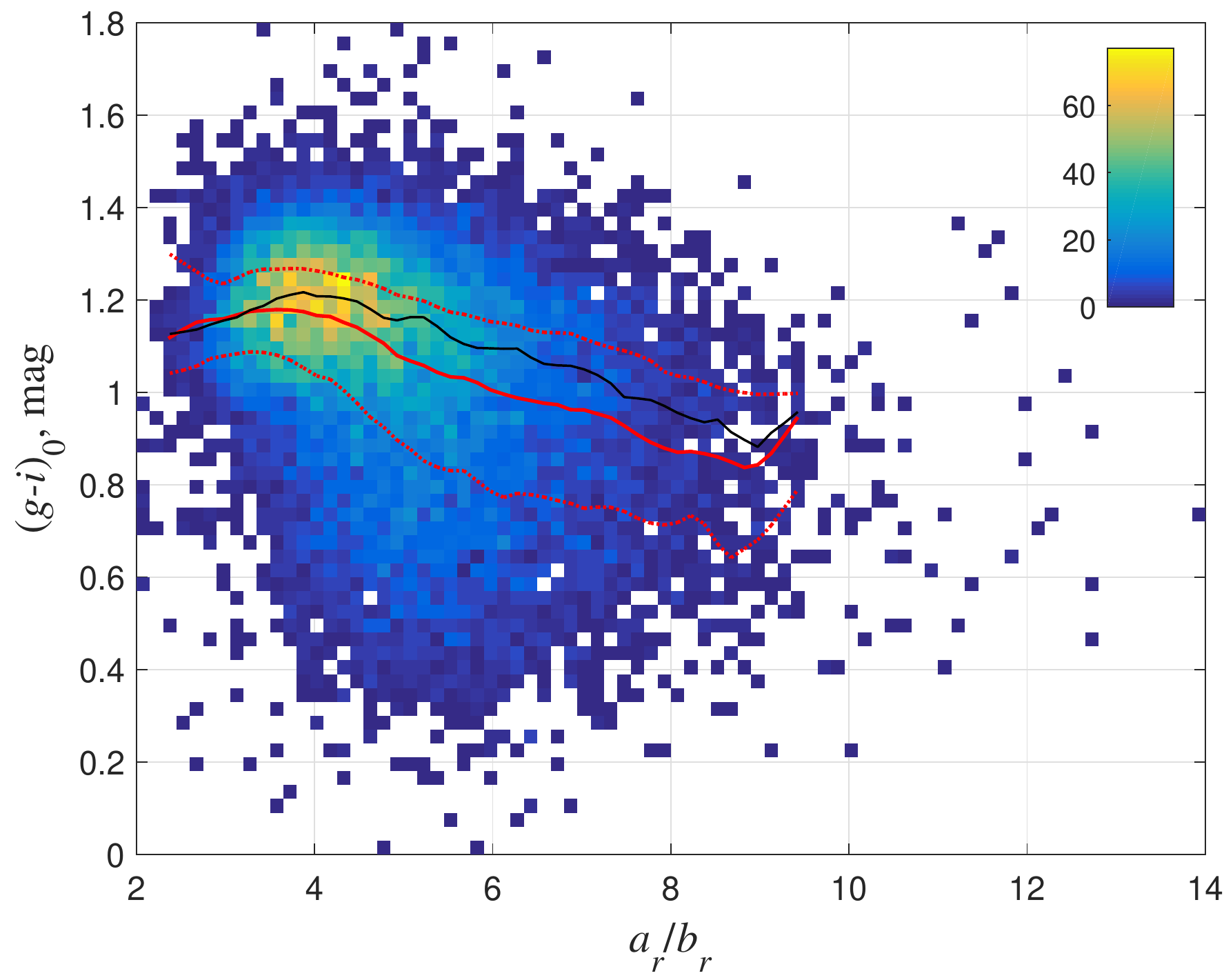}
\caption{
The colour $(g-i)_0$ versus inverse thickness $(a/b)_r$ for the EGIPS galaxies.
The galaxy colours are corrected for Galactic reddening.
The galaxy counts per bin are colour-coded according to the legend.
The running median is shown by the red line, while the 25 and 75 per cent quartiles are given by the red dot lines.
The black line represents the running mode of the distribution.
}
\label{fig:color+axisRatio}
\end{figure}

Despite the quite wide spread in colours, 
we see a clear trend of the galaxy colour $(g-i)_0$ on the galaxy thickness (see Fig.~\ref{fig:color+axisRatio}).
Thick galaxies form a cloud with a typical colour of $(g-i)_0\approx1.2$~mag,
which is systematically redder than the distribution of thin galaxies.
The colour of the distribution maximum is almost constant for galaxies with an axis ratio in the range from 3 to 5.
This behaviour of red galaxies was reported earlier by~\citet{2014ApJ...787...24B} and \citet{2009PASP..121.1297K}.
The galaxies with $(a/b)_r>5$ turn out to be bluer on average as the axes ratio increases.
The `comet' tail of the distribution shows similar trends
$(g-i)_0\propto(-0.055\pm0.004) (a/b)_r$ and $\propto(-0.066\pm0.005) (a/b)_r$
for the running median and the running mode, respectively.
The colour reflects the variation in the stellar populations and the recent star-formation history of disc galaxies.
It is obvious that the thinnest galaxies have, on average, younger stellar populations.

The dependence of the galaxy thickness on morphology has been discussed for a long time.
\citet{1972MmRAS..75...85H} found that 
the largest known axis $a/b$ ratios for galaxies of different morphologies
increase smoothly from the elliptical to Sd-type galaxies with sharp drop for irregular galaxies.

Based on a modest sample of edge-on galaxies, \citet{1998MNRAS.299..595D} found that 
galaxies become systematically thinner when going from S0 to Sc,
while Sd are likely to be thicker than Sc galaxies.  
However, based on the accurate 2D/3D decomposition of galaxy images for a representative sample of edge-on galaxies, 
\citet{2015MNRAS.451.2376M} reconsidered the dependence between the morphological type and the disc flatness 
and did not find a significant correlation between these two galaxy quantities. 
Nor did they find a dependence between the bulge-to-total luminosity ratio 
(one of the quantitative characteristics for the Hubble classification) and the disc flatness. 
The correlation between the galaxy colour and disc flatness appeared poor as well. 
In addition to that, \citet[see Fig.~7]{2014ApJ...787...24B} did not find a significant trend between the disc thickness 
and the overall galaxy $(g-r)_0$ colour.
Since we did not perform bulge-disc decomposition in this study,
the relation between the galaxy colour and the \textit{disc} axis ratio
(similar to what is done for the axis ratio of the \textit{galaxy as a whole} in Fig.~\ref{fig:color+axisRatio}) 
deserves special attention in a future study.

Note that superthin galaxies, $(a/b)_g>10$, indicated with pink dots in Fig.~\ref{fig:galaxyCMD+axisRatio}, avoid the red sequence.
Their distribution does not differ from that of other spiral galaxies, showing the same colours and absolute magnitudes.
The good consistency of the distribution of superthin galaxies with the rest of the blue edge-on spirals 
also indicates the same effect of dust in both normal and superthin galaxies.

\begin{figure}
\centering
\includegraphics[width=0.45\textwidth]{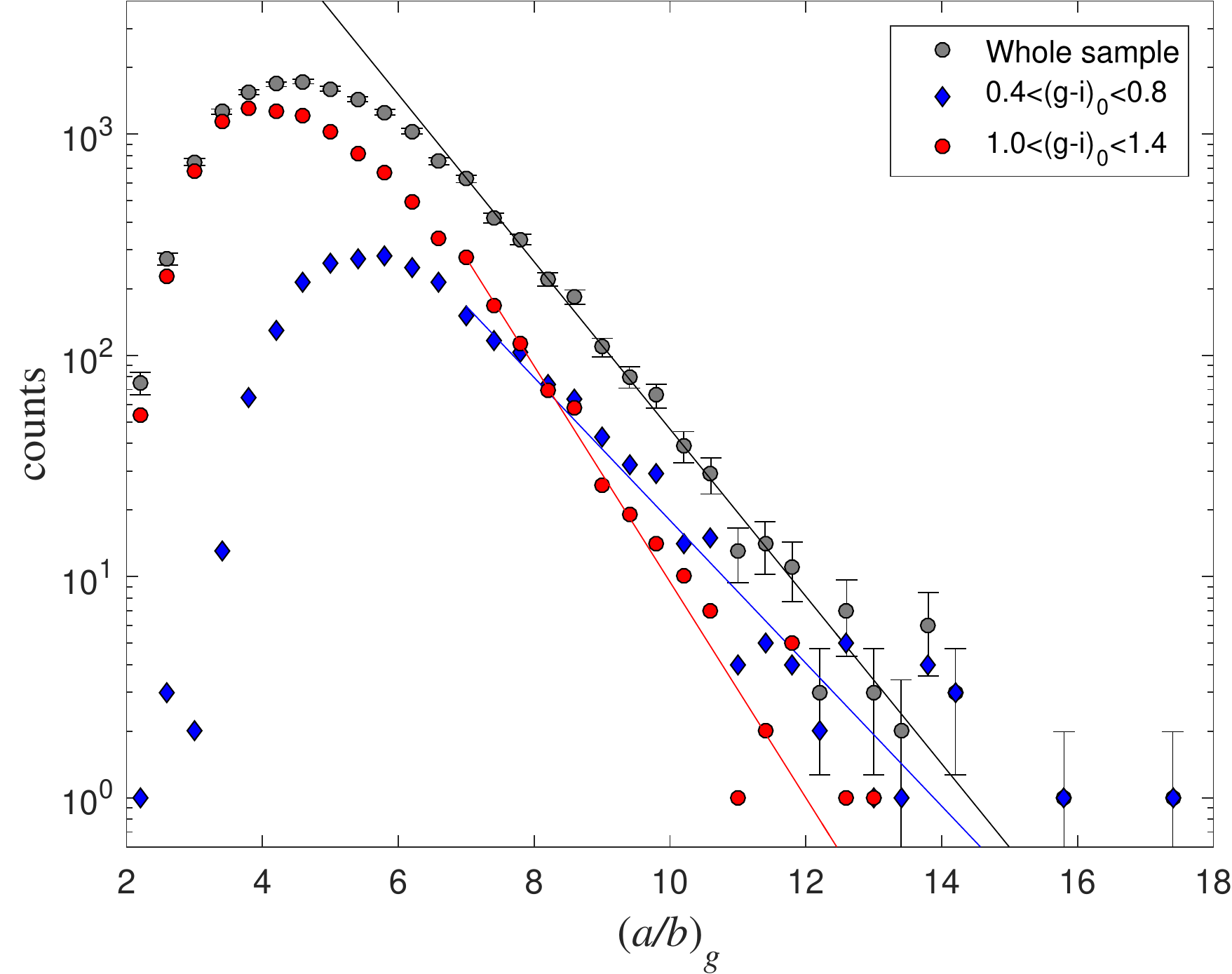}
\caption{
Distribution of the EGIPS galaxies by the axis ratio in the $g$ band.
The $(g-i)_0$ colour-selected subsamples are shown with different symbols and different colours.
}
\label{fig:AxisRatio}
\end{figure}

The minimal possible thickness of a stellar disc is directly related to its dynamic stability
and is regulated by the mass of the surrounding dark matter halo~\citep{1976AJ.....81...30H, 1991SvAL...17..374Z, 2002AstL...28..527Z, 2006AstL...32..649S,2010MNRAS.401..559M}.
The stellar disc at the stability limit
should obey the relationship $z_0/h\sim M_d/M_t$ \citep{2002AstL...28..527Z}
between the vertical-to-radial scale-length ratio of the disc, $z_0/h$, 
and the relative contribution of the disc-to-total galaxy mass, $M_d/M_t$.
From this it follows that the larger the relative mass of the dark matter halo, 
the smaller the relative thickness of the disc can be.
To find the smallest possible true (inclination corrected) relative thickness,
\citet{1994AstL...20....8K} proposed to reconstruct the density distribution
of the true axis ratio of galaxies from the distribution function of the observed ratios solving an integral equation.
They defined the maximal true $a/b$ axis ratio as such 
that the density distribution of the sample is 1 for the given axis ratio~\citep[][equation 6]{1994AstL...20....8K}.
In our case, the observed axis ratio distribution of flat, $a/b>7$, EGIPS galaxies is nearly exponential, 
$\log N \propto (-0.88\pm0.03)(a/b)_g$, with a slight excess for the thinnest galaxies (Fig.~\ref{fig:AxisRatio}).
Following the methodology by \citet{1994AstL...20....8K}, 
we estimated the maximum true axis ratio of galaxies in the EGIPS catalogue to be equal to $(a/b)_g=19.9$.
Taking into account the relations between EGIPS and RFGC sizes (see Section~\ref{sec:Comparison}),
this corresponds to the $a/b=25.7$ in the RFGC diameter system.
This value is actually the same as the peak true axis ratio 25.8 obtained by \citet{1994AstL...20....8K} 
for the sample of the RFGC galaxies.

\begin{figure}
\centering
\includegraphics[width=0.45\textwidth]{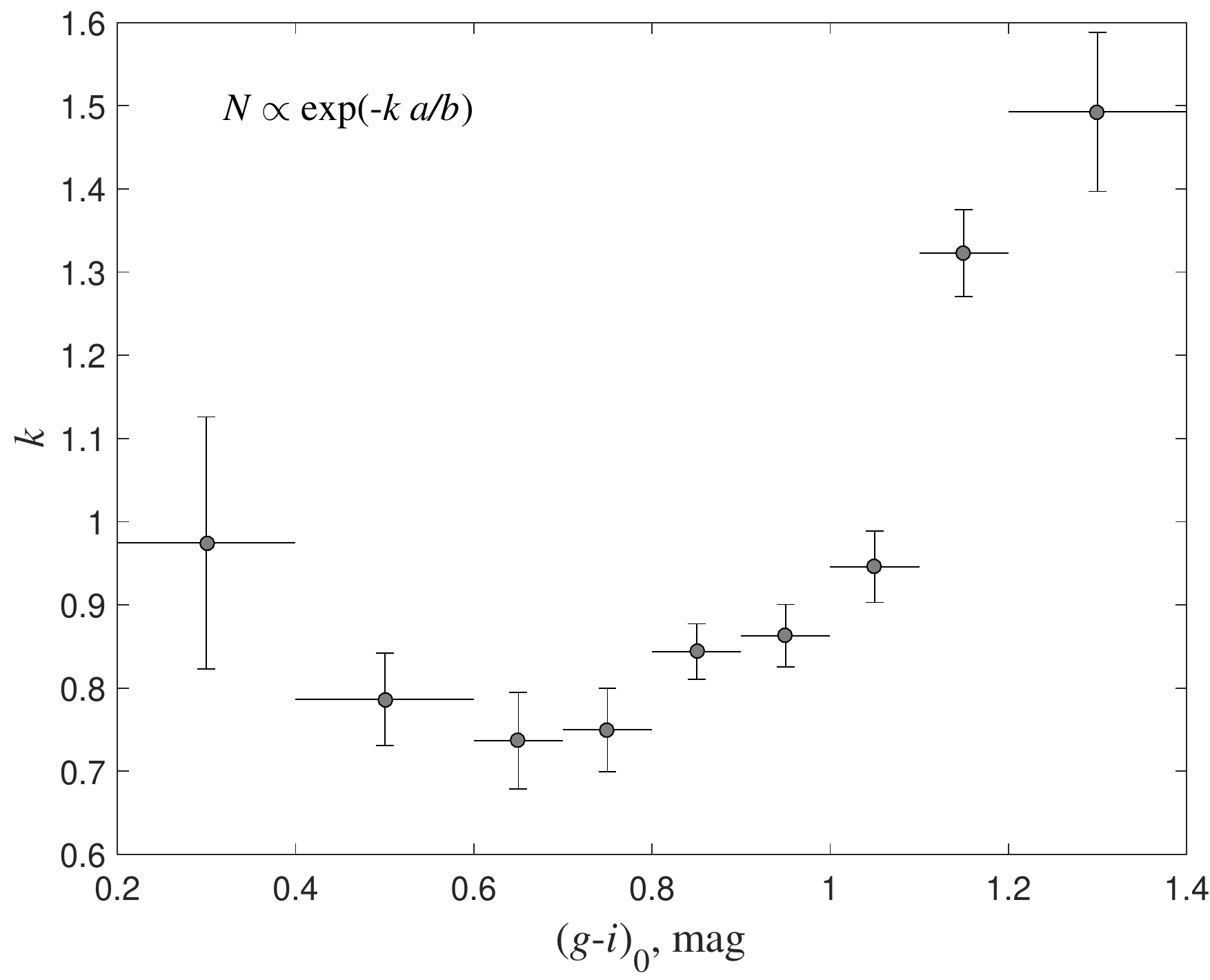}
\caption{
Dependence of the rate of the exponential decline for the distribution function of the axis ratios on colour.
Each value is calculated for a colour range indicated by the horizontal lines.
}
\label{fig:ColorAxisRatio}
\end{figure}

Figure~\ref{fig:AxisRatio} shows that the axis ratio functions
of red, $1.0\le(g-i)_0\le1.4$, and blue, $0.4\le(g-i)_0\le0.8$, galaxies are different.
The number of the red galaxies drops with increasing $a/b$ much faster than for the blue galaxies.
This is reflected in the fact that the early-type galaxies are systematically thicker than the late-type ones.
The effect is also seen in Fig.~\ref{fig:ColorAxisRatio},
which shows the dependence of the rate of the exponential decline, $k$, on the galaxy colour,
where the distribution of the axis ratios (in the $g$ band) for flat, $a/b>7$, galaxies 
is approximated by an exponential function $N\propto\exp(-k a/b)$ in narrow ranges of the $(g-i)_0$ colour.
The figure reveals a sharp change in the distribution around a colour of $(g-i)_0\approx1.1$~mag.
The value of $k$ gradually decreases for bluer galaxies in the colour range from 0.6 to 1.1.
This behavior has a natural explanation:
the youngest stellar populations form the thinnest subsystems in the galaxy discs.
\citet{1994AstL...20....8K} found a similar trend analyzing a sample of 4455 flat galaxies.
The rate of decline becomes gradually flatter in the transition from Sb to Sd galaxies.
However, the distribution of the bluest $(g-i)_0\le0.6$ EGIPS galaxies
unexpectedly changes the behaviour and falls steeper.
This behavior is expected for dwarf galaxies, which are systematically thicker~\citep{2013MNRAS.436L.104R},
but our sample contains too few dwarf galaxies to trace this effect robustly.

\section{Conclusions}

In this study we have presented the largest catalogue of edge-on galaxies to date. 
The sample is constituted by 16,551 objects found in publicly available Pan-STARRS images.
The public access to the EGIPS catalogue is supported by the Edge-on Galaxy Database\footnote{\url{https://www.sao.ru/edgeon/}} \citep{2021AstBu..76..218M}.

In this project, we have intensively used a convolutional neural network.
The catalogue of genuine edge-on galaxies EGIS~\citep{2014ApJ...787...24B} 
was used as a training sample for the CNN.
It allowed us to significantly improve the quality of the candidate selection.
Finally, all candidates were visually verified by professional astronomers
to screen out image artefacts, wrong classifications, and non-edge-on galaxies.

For all the galaxies in the catalogue, we performed \textsc{SExtractor} photometry 
and estimated non-parametric morphological quantities using \textsc{statmorph}.
A comparison with aperture photometry based on SDSS images showed the reliability of our data 
for galaxies in the 13.8--17.4 $r$-band magnitude range.
Our sample of edge-on galaxies is complete for galaxies with $a\ge6$~arcsec,
where $a$ is the standard deviation of the light distribution along the major axis of the galaxy
as defined in \textsc{SExtractor}.
We found a tight correlation between the disc scale-lengths from EGIS with the respective \textsc{SExtractor} sizes derived in this study.

The cross-identification of the sample objects with the HyperLeda database~\citep{2014A&A...570A..13M}
has shown that over 63 per cent of our galaxies have redshift measurements.
It allowed us to estimate the effective depth of the catalogue to be approximately 12,000~km\,s$^{-1}$ in the CMB frame of reference.

One of the purposes of the creation of this catalogue is studying the scaling relationships for galactic stellar discs and bulges.
Our analysis revealed that:
\begin{itemize}
\item The sample of our galaxies shows a clear separation into the red sequence and blue cloud galaxy populations.
\item The galaxy thickness distribution varies with the galaxy colour:
\begin{itemize}
\item the red sequence galaxies are thicker than the blue cloud galaxies;
\item in the blue cloud, thinner galaxies are systematically bluer.
\end{itemize}
\item The axis ratio distribution declines with increasing galaxy axis ratio faster for the red galaxies than for the blue ones.
\item However, for the bluest edge-on galaxies, the decline in the distribution function increases again.
\item The edge-on galaxies are systematically redder than the general population of galaxies seen at arbitrary angles,
which is apparently related to the influence of the internal extinction in the galaxies.
\end{itemize}

\section*{Acknowledgements}
We thank the anonymous referee for her/his kind and helpful comments.

This research was supported by the Russian Science Foundation grant \textnumero~19--12--00145.
The work on the Edge-on Galaxy Database was supported by the Russian Foundation for Basic Research grant  \textnumero~19--32--90244.

The Pan-STARRS1 Surveys (PS1) and the PS1 public science archive have been made possible through contributions by 
the Institute for Astronomy, the University of Hawaii, the Pan-STARRS Project Office, the Max-Planck Society and its participating institutes, 
the Max Planck Institute for Astronomy, Heidelberg and the Max Planck Institute for Extraterrestrial Physics, Garching, 
The Johns Hopkins University, Durham University, the University of Edinburgh, the Queen's University Belfast, 
the Harvard-Smithsonian Center for Astrophysics, the Las Cumbres Observatory Global Telescope Network Incorporated, 
the National Central University of Taiwan, the Space Telescope Science Institute, 
the National Aeronautics and Space Administration under Grant No. NNX08AR22G 
issued through the Planetary Science Division of the NASA Science Mission Directorate, 
the National Science Foundation Grant No. AST-1238877, 
the University of Maryland, Eotvos Lorand University (ELTE), the Los Alamos National Laboratory, and the Gordon and Betty Moore Foundation.

This research has made use of `Aladin sky atlas' developed at CDS, Strasbourg Observatory, France.

\section*{Data availability}

The data underlying this article are available at the Edge-on Galaxy Database \url{https://www.sao.ru/edgeon/}.

\bibliographystyle{mnras}
\bibliography{ref}

\bsp	
\label{lastpage}
\end{document}